\documentclass[conference]{IEEEtran}
\IEEEoverridecommandlockouts
\usepackage[T1]{fontenc}
\usepackage{iftex}
\ifXeTeX
    \usepackage{newunicodechar}
    \newunicodechar{^^^^2014}{\textemdash}
\fi
\usepackage{cite}
\usepackage{amsmath,amssymb,amsfonts}
\usepackage{algorithmic}
\usepackage{graphicx}
\usepackage{textcomp}
\usepackage{xcolor}
\usepackage{url}
\usepackage{booktabs}
\usepackage{multicol}
\usepackage{multirow}
\usepackage{diagbox}
\usepackage{siunitx}
\usepackage{enumitem}
\usepackage{longtable}
\usepackage{array}
\usepackage{tabularx}
\usepackage{listings}
\usepackage{tcolorbox}
\tcbuselibrary{listings,skins,breakable}

\def\BibTeX{{\rm B\kern-.05em{\sc i\kern-.025em b}\kern-.08em
    T\kern-.1667em\lower.7ex\hbox{E}\kern-.125emX}}

\lstdefinestyle{yamlstyle}{
basicstyle=\ttfamily\scriptsize,
breaklines=true,
breakautoindent=true,
breakindent=12em,
showstringspaces=false,
numbers=left,
numberstyle=\tiny\color{gray},
tabsize=2
}

\newtcblisting{errorlisting}[2][]{
listing only,
breakable,
enhanced,
colback=red!3,
colframe=red!50,
coltitle=white,
fonttitle=\scriptsize\bfseries,
title=#2,
listing options={style=yamlstyle,#1}
}

\newcommand{\paperinput}[1]{%
  \IfFileExists{latex/#1.tex}{\input{latex/#1}}{\input{#1}}%
}

\newtcolorbox{userbox}[1]{
    colback=blue!5!white,
    colframe=blue!75!black,
    fontupper=\small,
    fonttitle=\small\bfseries,
    title={Human User (#1)},
    enhanced,
    breakable,
    attach boxed title to top left={yshift=-2mm, xshift=2mm},
    boxed title style={colback=blue!75!black}
}

\newtcolorbox{agentbox}[1]{
    colback=green!5!white,
    colframe=green!60!black,
    fontupper=\small,
    fonttitle=\small\bfseries,
    title={#1},
    enhanced,
    breakable,
    attach boxed title to top left={yshift=-2mm, xshift=2mm},
    boxed title style={colback=green!60!black}
}

\newtcolorbox{systembox}[1]{
    colback=orange!5!white,
    colframe=orange!70!black,
    fontupper=\small,
    fonttitle=\small\bfseries,
    title={#1},
    enhanced,
    breakable,
    attach boxed title to top left={yshift=-2mm, xshift=2mm},
    boxed title style={colback=orange!70!black}
}

\begin{document}

\title{Agent4Spec: An LLM-based Agentic System to Optimize API Specifications through Automated Validation and Market Analysis}

\author{
\IEEEauthorblockN{
Yongjian Tang\IEEEauthorrefmark{1}\IEEEauthorrefmark{2},
Costel Moraru\IEEEauthorrefmark{1},
Doruk Tuncel\IEEEauthorrefmark{1},
Hanya Elhashemy\IEEEauthorrefmark{1},
Steffen Klepke\IEEEauthorrefmark{1},
Thomas Runkler\IEEEauthorrefmark{1}\IEEEauthorrefmark{2}
}
\IEEEauthorblockA{
\IEEEauthorrefmark{1}Siemens Research (RPD), Munich, Germany\\
\IEEEauthorrefmark{2}Technical University of Munich, Munich, Germany
}
}

\maketitle

\begin{abstract}
API specifications serve as critical documentation for modern software development, directly impacting integration efficiency and client code generation.
While recent advancements have leveraged Large Language Models (LLMs) to generate initial specifications from source code, existing approaches treat this process as a one-time event -- overlooking the needs for continuous validation, long-term maintenance, and market-driven evolution of these artifacts.
To address this issue, we present Agent4Spec, an autonomous multi-agent framework featuring two distinct pipelines: (1) a validation pipeline that detects and self-corrects OpenAPI syntax errors, and (2) a market analysis pipeline that conducts competitive research, identifies functional gaps against industry leaders, and automatically generates missing API-spec functionalities.
We evaluate Agent4Spec on two curated datasets of API specifications spanning both general-purpose and industrial scenarios.
Among 9 SOTA LLMs, the top-performer achieves 100\% syntax error detection rates and over 95\% implementation success for proposed corrections, while securing over 50\% adoption rates for suggested endpoint improvements, representing a practical tool for API development.
\end{abstract}


\section{Introduction}

API specifications define the operational contract between service providers and consumers.
For REST services, OpenAPI Specification (OAS) documents describe endpoints, parameters, schemas, and error cases in a machine-readable form.
Such specifications support client integration, API testing, and long-term maintenance, while inaccurate or incomplete specifications can mislead developers and propagate failures across downstream tools.
As software ecosystems evolve, maintaining both the syntactic validity and practical relevance of these specifications becomes increasingly difficult.

Recent work has made substantial progress in generating API specifications from source code \cite{chen2026openaiopenapiautomatedgeneration,10.1145/3810241,tang2024fsponer,10.1145/3597503.3639137} and online documentation \cite{lazar-etal-2025-generating,lazar2024specrawler}.
Other studies have examined LLMs for API documentation understanding and updates \cite{chen2025llmsmeetapidocumentation,11185874,tang2025fewshot,li2025empirical}.
However, these efforts mainly target initial specification construction or document-level maintenance.
They provide limited support for the post-generation stage, where developers must validate full OAS files, implement repairs safely, and decide whether the API should evolve to match current market expectations.

\begin{figure*}
\centering
\includegraphics[width=1\linewidth]{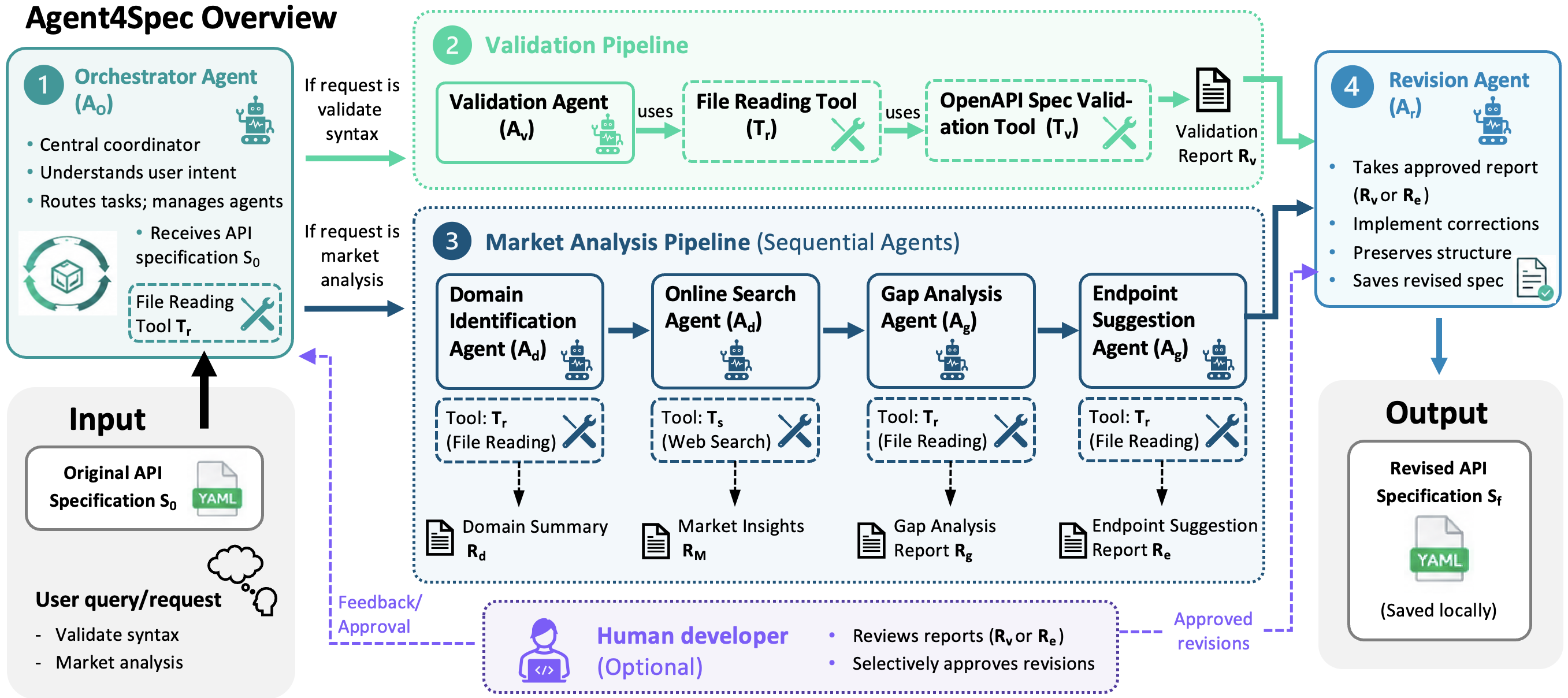}
\caption{Agent4Spec consists of four core components: an orchestrator agent $A_o$ acting as the central controller and communicator; a validation agent $A_v$ to detect and resolve specification syntax errors; multiple sequential agents $A_d, A_m, A_g, A_e$ to explore current technology trends and identify gaps against market leaders for improvement; and finally a revision agent $A_r$ that automatically updates specifications based on developer-approved suggestions.}
\label{agent_overview}
\end{figure*}

To address this gap, we present Agent4Spec, a multi-agent system for post-stage API specification validation and enhancement.
As shown in Figure \ref{agent_overview}, Agent4Spec uses an orchestrator agent to route developer requests into two task-specific pipelines.
The validation pipeline detects OAS syntax errors and proposes corrections through tool-assisted checking.
The market analysis pipeline identifies the API domain, searches for competing services and current trends, analyzes functional gaps, and suggests endpoint-level improvements.
A revision agent can then apply approved changes to the original specification, with optional human review before market-driven changes are implemented.

We evaluate Agent4Spec on two curated datasets, Guru-Spec and Industry-Spec, covering both general-purpose and industrial API scenarios.
The experiments include nine LLMs across proprietary frontier and compact open-weight tiers, allowing us to study both effectiveness and resource trade-offs.
Our main contributions can be summarized as follows:

\begin{itemize}
    \item [(i)] We formulate API specification enhancement as a post-generation workflow that jointly considers syntax validation, market analysis, and specification revision.
    \item [(ii)] We propose Agent4Spec, a multi-agent framework that coordinates specialized agents for validation, competitive analysis, endpoint suggestion, and optional human-approved revision.
    \item [(iii)] We construct two evaluation datasets, Guru-Spec and Industry-Spec, to measure Agent4Spec across general-purpose and industrial API specifications.
    \item [(iv)] We conduct an extensive evaluation across 9 LLMs. The strongest frontier models achieve 100\% syntax-error detection, over 95\% correct implementation for proposed corrections, and up to 1.82 human-approved endpoint suggestions out of three, while the same evaluation also exposes practical limitations of compact models in tool use, full-file revision, and market-oriented analysis.
\end{itemize}

\section{Related Work}

\subsection{API Specification Generation}
API specification generation has been studied from both implementation artifacts and public documentation.
For source-code-based generation, Respector \cite{10.1145/3597503.3639137} uses static and symbolic program analysis to infer endpoint methods, parameters, responses, and constraints from REST API implementations.
Recent LLM-based approaches extend this direction: LRASGen \cite{10.1145/3810241} generates RESTful API specifications from cleaned source code, while OOPS \cite{chen2026openaiopenapiautomatedgeneration} models cross-file API dependencies and applies self-refinement to improve OpenAPI Specification (OAS) compliance.
In parallel, documentation-based systems such as SpeCrawler \cite{lazar2024specrawler} and OASBuilder \cite{lazar-etal-2025-generating} generate OAS files from online API documentation through scraping and LLM-based synthesis.
These methods reduce the cost of producing initial specifications, but they mainly assume a specific generation setting.
Agent4Spec instead targets the post-generation stage, where an existing OAS file must be validated, repaired, and potentially enhanced.

\subsection{API Documentation Understanding and Maintenance}
Several studies examine whether LLMs can understand or update API documentation.
Li et al. \cite{li2025empirical} evaluate LLM accuracy in API documentation understanding across programming languages, and Chen et al. \cite{chen2025llmsmeetapidocumentation} study retrieval-augmented API documentation for code generation.
Lee et al. \cite{11185874} move closer to maintenance by extracting API change rules from repository histories and prompting LLMs to update documentation snippets.
These works show that LLMs can support API-documentation tasks, but they focus on textual documentation or downstream code assistance.
In comparison, Agent4Spec focuses on machine-readable OAS artifacts and couples validation reports with full-specification revision.

\subsection{Agentic Workflows and Evaluation}
LLM-based agentic systems have been widely applied to complex software engineering tasks \cite{tang2026llmbasedagenticsystemssoftware}.
ChatDev \cite{qian2024chatdev}, Agent4RE \cite{tang2026agent4re}, and MetaGPT \cite{hong2024metagpt} organize multiple agents into role-based software-development workflows, while SWE-agent \cite{yang2024sweagent}, Agent4cs \cite{tang2026agent4cs}, and AutoCodeRover \cite{zhang2024autocoderover} focus on repository-level issue resolution and program improvement.
These systems demonstrate the value of task decomposition, tool use, and iterative feedback for software development.
However, they mainly operate on source code, repositories, or natural-language development tasks.
We propose Agent4Spec to target machine-readable OAS maintenance, where the output must remain syntactically valid, structurally consistent, and controllable by developers before endpoint-level changes are applied.

\section{Agent4Spec}

As illustrated in Figure \ref{agent_overview}, Agent4Spec starts with an orchestrator agent $A_o$, which serves as the central coordinator -- interacting with the developers while controlling over all other specialized sub-agents.
Once an API specification $S_0$ is received, the orchestrator directs the workflow based on the user's query -- either triggering the validation agent to detect syntax errors in the document, or forwarding it to a series of market research agents to identify current technology trends and leading API service competitors.
Notably, only one of these two pathways is executed before calling the revision agent, ensuring that the system is not overwhelmed with excessive revision instructions.
Optionally, human developers can be incorporated into this workflow before the revision agent is triggered, reviewing and selectively approving the suggestions produced by either the validation or market analysis pipelines.

\subsection{Validation Pipeline}
We equip the validation agent with a file reading tool $T_r$ and an OpenAPI specification validation tool $T_v$.
Given a local file path, $T_r$ enables the agent to thoroughly parse and examine the uploaded API YAML file in detail, while $T_v$ utilizes the Python library \texttt{openapi\_spec\_validator} to execute the parsed YAML file and detect whether any syntax errors exist.
Based on the resulting error message, the validation agent proposes fixing solutions to resolve the identified issues.
Using the notation in Table \ref{tab:notation}, the agent output can be defined as:
\begin{equation}
R_v = A_v( T_v(T_r(S_0)) ).
\end{equation}
In the next step, the revision agent can be invoked to implement the proposed solutions automatically.

\begin{table}[htbp]
\caption{Notation used in Agent4Spec.}
\label{tab:notation}
\centering
\resizebox{1\linewidth}{!}{
\setlength{\tabcolsep}{4pt}
\begin{tabular}{ll}
\toprule
\textbf{Notation} & \textbf{Description / Functional Definition} \\
\midrule

$S_0$ & Original OpenAPI specification provided by the user \\
$S_f$ & Final revised OpenAPI specification \\
$R_v$ & Report of validation errors with proposed correction suggestions\\
$R_d$ & Identified API domain and domain summary \\
$R_m$ & Synthesized market analysis report \\
$R_g$ & Gap analysis report with prioritized suggestions\\
$R_e$ & Endpoint-level report highlighting improvement suggestions \\
$R$ & Intermediate report from the validation pipeline or the market analysis pipeline \\
$R'$ & Revisions approved by human developers in the pipeline report \\
\midrule

$T_r(S)$ & \textbf{File Reading Tool:} Parses and loads an OpenAPI specification \\
$T_v(S)$ & \textbf{Validation Tool:} Checks OpenAPI compliance and detects syntax errors \\
$T_s(R_d)$ & \textbf{Web Search Tool:} Retrieves market information for API domain $R_d$ \\

\midrule

$A_o(S_0)$ & \textbf{Orchestrator Agent:} Routes workflow execution and coordinates agents \\
$A_v(S_0)$ & \textbf{Validation Agent:} Suggest corrections from validation errors \\
$A_d(S_0)$ & \textbf{Domain Identification Agent:} Determines API domain and key characteristics \\
$A_m(R_d)$ & \textbf{Market Analysis Agent:} Searches and synthesizes collected market information \\
$A_g(R_m,S_0)$ & \textbf{Gap Analysis Agent:} Identifies the gap between our API service \& market leaders \\
$A_e(R_g,S_0)$ & \textbf{Endpoint Suggestion Agent:} Generates endpoint-level improvement proposals \\
$A_r(R,S_0)$ & \textbf{Revision Agent:} Revises the specification using approved suggestions \\
\bottomrule
\end{tabular}
}

\end{table}

\subsection{Market Analysis Pipeline}
Alternative to syntax validation, orchestrator can also initiate a market analysis pipeline upon user requests, which consists of four sequential agents.

\begin{enumerate}
    \item \textbf{Domain Identification Agent:} The first agent $A_d$ reads through the uploaded API specification, determines the API domain, and summarizes its key developments as bullet points:
    \begin{equation}
        R_d = A_d(T_r(S_0)).
    \end{equation}

    \item \textbf{Online Search Agent:} The following agent $A_m$ conducts comprehensive market analysis by leveraging an online search tool $T_s$.
    Given the specified API domain, it formulates search queries, executes online web searches, and selects maximum 3 results per query, prioritizing authoritative sources such as vendor documentation and reputable technology media.
    Together with the browsed website links, the agent synthesizes all retrieved information into a market summary:
    \begin{equation}
        R_m = A_m(T_s(R_d)),
    \end{equation}
    which delivers structured insights on market leaders, landscape, emerging trends, and pricing models with quantified statistics where available.

    \item \textbf{Gap Analysis Agent:} The gap analysis agent $A_g$ compares the user's API service against the identified market leaders in $R_m$:
    \begin{equation}
        R_g = A_g(R_m,T_r(S_0)).
    \end{equation}
    For each uploaded specification $S_0$, it highlights missing functionalities, design inconsistencies, ecosystem limitations, and potential competitive disadvantages.
    Findings are categorized into high-, medium-, and low-priority recommendations accompanied by technical impact assessments.

    \item \textbf{Endpoint Suggestion Agent:} The endpoint suggestion agent $A_e$ generates concrete API endpoints to address the prioritized gaps:
    \begin{equation}
        R_e = A_e(R_g,T_r(S_0)),
    \end{equation}
    which contains endpoint-level enhancements, API schemas, migration guidance, along with summarized market insights.
\end{enumerate}

\subsection{Specification Revision Agent}
Given either validation reports $R_v$ or market-oriented endpoints suggestion $R_e$ from previous agents, the revision agent $A_r$ implements the corrections to specification $S_0$ while preserving its content structure.
Furthermore, we develop an iterative re-validation mechanism to improve the accuracy in large specifications, where the revised documents are repeatedly sent back to the validation agent for $n$ rounds until all errors are eliminated.
Formally, the final specification $S_f$ is:
\begin{equation}
\begin{gathered}
S_f = A_r^{(n)}(R,T_r(S_0)),\\[1ex]
\scalebox{0.95}{$
R = \begin{cases}
R_v, & \text{validation mode}\\
R_e, & \text{market analysis mode}
\end{cases}$}.
\end{gathered}
\end{equation}
Once completed, the agent saves the corrected specification locally with appropriate file naming, and transfers control back to the orchestrator, ensuring seamless workflow integration within the system.

\subsection{Human-in-the-Loop}
Based on the improvement suggestions from the two pipelines, human developers can optionally be involved to review and approve these recommendations before handing over to the revision agent.
The finalized report, after human approval, can be represented as $R' = H(R)$.
This step allows flexible quality control in practice and embeds expert feedback into critical decision-making processes.

\section{Experimental Setup}

\subsection{Datasets}
To evaluate Agent4Spec across both general-purpose and industrial API ecosystems, we have curated two datasets of OpenAPI specifications with complementary features: 1) \textbf{Guru-Spec} contains 300 publicly available API specifications collected from APIs.guru \cite{api_guru}, covering diverse domains such as cloud services, finance, e-commerce, and transportation.
2) In contrast, \textbf{Industry-Spec} focuses on 200 industrial APIs obtained from Xcelerator Developer \cite{developer_portal}.
These APIs target industrial applications and enterprise services, covering domains such as automation, Internet of Things, asset management, and manufacturing.
Together, these two datasets enable evaluation across both broad web services and industrial scenarios.
Detailed statistics are presented in Table \ref{tab:dataset_stats}.

\begin{table}[htbp]
\caption{Statistics of the curated datasets.}
\label{tab:dataset_stats}
\centering
\resizebox{1\linewidth}{!}{
\setlength{\tabcolsep}{12pt}
\begin{tabular}{l S[table-format=7.1] S[table-format=7.1]}
\toprule
\textbf{Statistic} & \textbf{Guru-Spec} & \textbf{Industry-Spec} \\
\midrule
Number of specifications & 300 & 200 \\
Total dataset tokens & 3483222 & 1684932 \\
Maximum tokens in a single spec & 59836 & 36145 \\
Average tokens per spec & 11610.7 & 8424.7 \\
Average lines per spec & 1549.2 & 934.4 \\
\bottomrule
\end{tabular}
}
\end{table}

\subsection{Models}
We strategically select 9 LLMs that span different performance tiers, as presented in Table \ref{tab:models}.
We deploy 5 frontier models via APIs, including Claude-Sonnet-4.6 \cite{claude_sonnet} from Anthropic, Gemini-3.1-Pro \cite{gemini31_pro} from Google, and GPT-5.2, GPT-4.1, GPT-4o \cite{gpt52,2024gpt41,hurst2024gpt4o} from OpenAI, which represent the forefront of LLM capabilities.
To complement these proprietary models, we incorporate four open-source alternatives that ensure reproducibility and broader applicability: Ministral3-8B \cite{mistral3}, LLaMA3.1-8B \cite{dubey2024llama} from Meta, Qwen3.5-9B \cite{qwen35} from Alibaba, and Gemma4-4B \cite{gemma4} from Google.
Large-scale language models are accessed through API gateways, while compact models are deployed via the open-source Ollama framework, with the temperature parameter of all models set to the minimal value.

\begin{table}[htbp]
\caption{Basic information of selected LLMs.}
\label{tab:models}
\centering
\scriptsize
\setlength{\tabcolsep}{4pt}
\resizebox{1\linewidth}{!}{%
\begin{tabular}{llllll}
\toprule
\textbf{Company} & \textbf{Model} & \textbf{Size} & \textbf{Context Window} & \textbf{Modalities (input)} & \textbf{Openness} \\
\midrule
\multirow{3}{*}{OpenAI} & GPT-5.2 & -- & 1M tokens & text/image & closed \\
& GPT-4.1 & -- & 1M tokens & text/image & closed \\
& GPT-4o & -- & 128k tokens & text/image/audio/video & closed \\
\midrule
\multirow{2}{*}{Google} & Gemini-3.1-Pro & -- & 1M tokens & text/image/audio/video & closed \\
& Gemma4-4B & 4B & 128k tokens & text/image/audio & open \\
\midrule
Anthropic & Claude-Sonnet-4.6 & -- & 1M tokens & text/image & closed \\
\midrule
Mistral & Ministral3-8B & 8B & 256k tokens & text/image & open \\
\midrule
Meta & LLaMA3.1-8B & 8B & 128k tokens & text-only & open \\
\midrule
Alibaba & Qwen3.5-9B & 9B & 1M tokens & text/image/video & open \\
\bottomrule
\end{tabular}%
}
\end{table}

\subsection{Metrics}
\label{ssubsection:metric}
We select an extensive set of evaluation metrics tailored to each task.
For syntax validation -- a task with verifiable outputs, we use quantitative metrics, including \textbf{1) error detection rate} (the percentage of errors identified by the system), \textbf{2) solution proposal rate} (the percentage of detected errors for which solutions are proposed), \textbf{3) solution correctness} (the percentage of proposed solutions that are accurate).
For API specification revision, we employ \textbf{1) correct implementation rate} (the percentage of solutions that are properly implemented), and \textbf{2) successful save rate} (the percentage of solutions locally saved).

Regarding market analysis, where no groundtruth answers are available, we apply LLM-as-a-judge \cite{llm_as_a_judge} complemented by human evaluation.
To be more specific, \textbf{1)} \textbf{LLM-as-a-judge} employs prompt-based assessment, with grades assigned on a scale from 1 to 5.
Given the empirical evidence that smaller LLMs often struggle in judging summary quality and even the best one frequently misjudges \cite{thakur-etal-2025-judging}, we only employ the most competent large-scale models -- Claude-Sonnet-4.6, Gemini-3.1-Pro, and GPT-5.2.
In comparison, \textbf{2) human evaluation} verifies the consistency of LLM-as-a-judge.
Three software developers assess summaries independently on a scale from 1 to 5, with final scores averaged for fairness.
We selected human evaluators based on their software development experience, code-review skills, and relevant domain expertise.
Finally, we chose two senior software engineers, each with three years of development experience, and one junior software developer in industry.
In addition to their own experience and judgment criteria, we also provided the prompt and grading schema used in LLM-as-a-judge to ensure consistent evaluation.
An example judge prompt to evaluate the quality of market analysis is demonstrated below:

\begin{tcolorbox}[colback=blue!5,colframe=blue!50,title=Market Research Evaluation --- LLM-as-a-Judge Prompt,fonttitle=\bfseries\footnotesize, boxrule=0.4pt,left=4pt,right=4pt,top=4pt,bottom=4pt,breakable]
\scriptsize
\ttfamily
You are an expert evaluator for API market research.
You will be given two inputs:\\
\textbullet\ An original API specification.\\
\textbullet\ A market summary produced by a web-based search agent.\\

\textbf{Your task:}\\
\textbullet\ Evaluate the market summary's semantic alignment with the API specification (scope, domain, capabilities, target users, ecosystem).\\
\textbullet\ Judge the intrinsic quality of the market summary itself (accuracy, sourcing, coverage, structure, clarity, relevance, currency).\\
\textbullet\ Provide one final rating (1--5) that balances both alignment and intrinsic quality.\\

\textbf{Rating Scale:}\\
1: Very Poor --- Major misalignment and low quality. Off-scope, factual errors, speculative claims, few/no credible sources, missing core market elements.\\
2: Poor --- Partial relevance with notable gaps. Some correct points but inconsistent, weak citations, outdated information, limited coverage of competitors/pricing/adoption; unclear or contradictory.\\
3: Fair --- Mostly aligned with moderate quality. Generally accurate, some useful insights, but gaps in coverage; organization acceptable but needs improvement.\\
4: Good --- Strong alignment and solid quality. Well-structured, relevant, mostly comprehensive with credible, recent sources; clear, concise synthesis and actionable takeaways; minor omissions or clarity issues.\\
5: Excellent --- High alignment and high intrinsic quality. Comprehensive, well-cited, and balanced; clearly connects market insights to the API's scope; precise, organized, actionable, no contradictions; helpful additions beyond the specification.\\

\textbf{Output Format (must follow exactly):}\\
Evaluation: (your comments and arguments for the rating)\\
Total rating: (a number between 1 and 5)\\

\textbf{Return Only:}\\
Evaluation: ...\\
Total rating: ...
\end{tcolorbox}

\section{Experiments -- Syntax Validation Pipeline}
Following the architecture in Figure \ref{agent_overview}, we first evaluate the validation pipeline based on the quantitative metrics defined in Section \ref{ssubsection:metric}.
For each collected API specification, we manually inject a syntax error to assess the system's performance, including invalid data types, missing parameters, and inconsistent naming.
Representative examples of injected errors are listed below.

\noindent\textbf{1) An indentation mismatch} can alter the YAML hierarchy, causing structural validation failures.
\begin{errorlisting}{Indentation Error in OpenAPI YAML}
paths:
  /users:
    get:
      responses:
        '200':
          content: 
            application/json:
            schema: # <- ERROR: indentation mismatch
\end{errorlisting}

\noindent\textbf{2) A specification-compliance error} may omit a mandatory response description field.
\begin{errorlisting}{Missing Required Response Description}
paths:
  /status:
    get:
      responses:
        '200': # <- ERROR: Missing required description
          content:
            application/json:
              schema:
\end{errorlisting}

\noindent\textbf{3) Type-mismatch errors} conflict with the declared property definitions.
\begin{errorlisting}{Mismatch between Type and Value}
components:
  schemas:
    UserConfiguration:
      type: object
      properties:
        retryLimit:
          type: integer    # <- ERROR: Declared as a numeric integer
          default: "three" # <- ERROR: String value conflicts with type
        isPremium:
          type: boolean
          example: 1       # <- ERROR: Integer used instead of true/false
\end{errorlisting}

\subsection{Syntax Error Detection}
Table \ref{tab:validation} compares Agent4Spec with a single-prompt baseline, where the same LLM is asked to handle validation in one prompt. 
Across all evaluated models and datasets, Agent4Spec outperformed the baseline in error detection, solution proposal, and solution correctness. 
When $T_v$ was invoked reliably, Agent4Spec detected nearly all syntax errors: Claude-Sonnet-4.6 and Gemini-3.1-Pro reached 100.0\% detection on both datasets, while GPT-5.2 and GPT-4.1 remained above 99.0\%.
GPT-4o was slightly lower, but still detected 96.0\% of errors on Guru-Spec and 94.5\% on Industry-Spec. 
The system proposed a repair for every error it detected, as reflected in the identical percentages for detected errors and proposed solutions.
Regarding the correctness of solutions, frontier models achieved 91.5--99.5\% correctness, while Qwen-3.5-9B and Ministral-3-8B dropped to 87.0\% and 87.5\% on Industry-Spec. The other two compact models LLaMA-3.1-8B and Gemma-4-4B showed a typical failure mode: they often stated that they would use $T_v$ but failed to activate it, yielding detection rates and correct-solution rates no higher than 8.0\%.

\subsection{Syntax Error Revision}
Table \ref{tab:revision} evaluates whether the single-prompt baseline and Agent4Spec can implement the proposed fixes and save a complete revised specification in its first attempt.
This task is more demanding than syntax-error detection because the revision agent must modify the full API specification file while preserving its original structure. Compared with the baseline, Agent4Spec generally achieved higher correct-implementation and save rates. Large-scale frontier models saved almost all revised specifications, with Claude-Sonnet-4.6, Gemini-3.1-Pro, GPT-5.2, and GPT-4.1 reaching 100.0\% save rates on both datasets. 
Their correct-implementation rates were lower, ranging from 82.0\% to 98.3\% on Guru-Spec and from 73.5\% to 98.5\% on Industry-Spec. This gap shows that 
saving and correctness are separate tool functions: an agent may save a revised specification even when the implemented solution is incorrect.
In comparison, compact open-weight models struggled in the full-spec revision setting.
Qwen-3.5-9B achieved only 10.3\% and 5.5\% correct implementation on Guru-Spec and Industry-Spec, while Ministral-3-8B, LLaMA-3.1-8B, and Gemma-4-4B were close to zero.
These results indicate that full-specification revision remains challenging for smaller models.
Iterative re-validation after each revision attempt is necessary to verify the quality of generated API specifications.

\section{Experiments -- Market Analysis Pipeline}
Alternatively, the market analysis pipeline can be initiated.
Since no standard answers exist for such open-ended tasks, we employ LLM-as-a-judge alongside human evaluation.

\subsection{Domain Identification}
As the initial step of the market analysis pipeline, domain identification requires the agent to infer the API scope and summarize relevant developments.
\begin{figure}[h]
\centering
\includegraphics[width=1\linewidth]{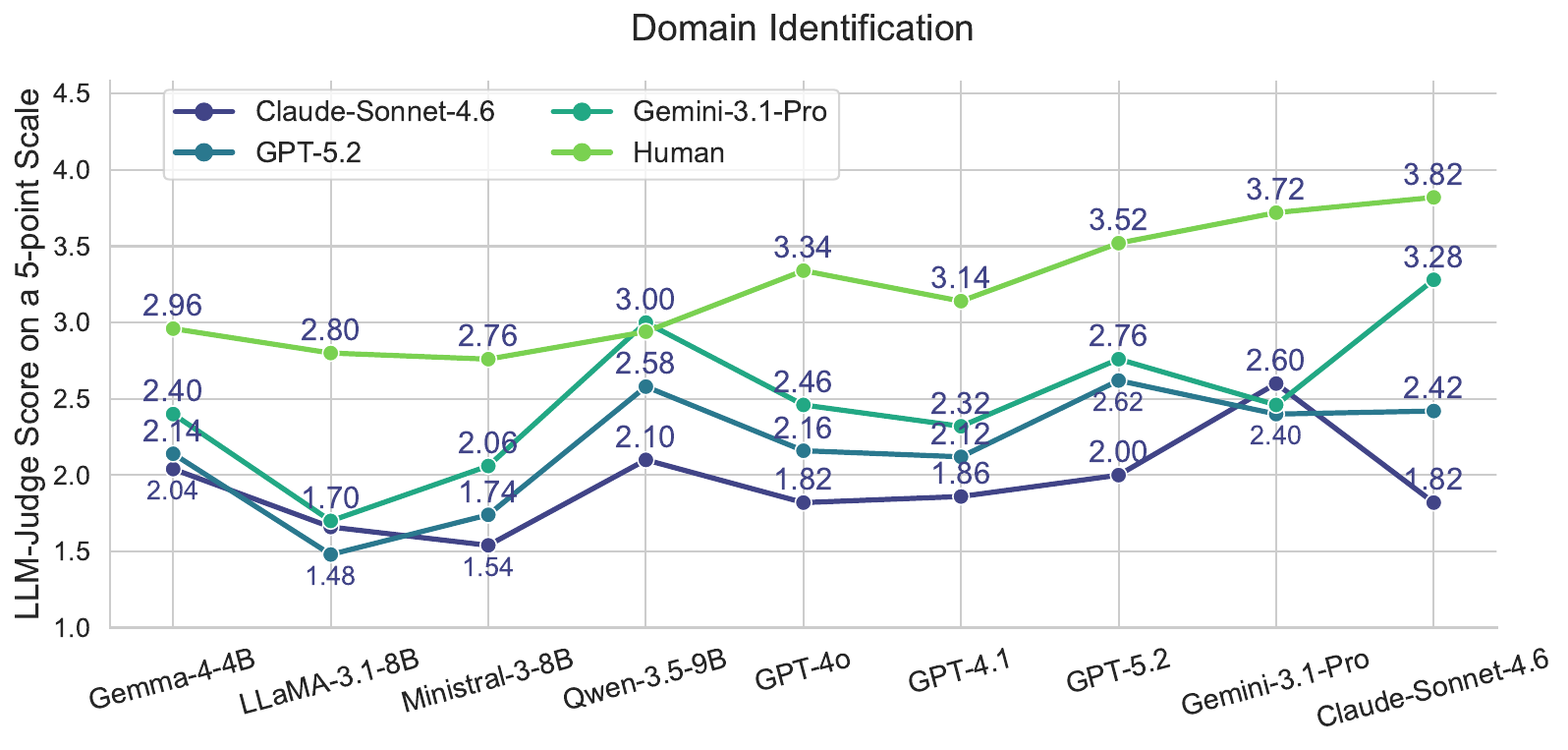}
\caption{Evaluation on domain identification using LLMs and human ratings.}
\label{fig:domain_identification}
\end{figure}
\begin{table*}[t]  
\centering  
\caption{Comparing the single-prompt baseline with Agent4Spec to detect syntax errors.}  
\label{tab:validation}  
\scriptsize  
\setlength{\tabcolsep}{4pt}  
\renewcommand{\arraystretch}{1.08}  
\resizebox{0.99\textwidth}{!}{%
\begin{tabular}{@{}ll|cccccc|cccccc@{}}  
\toprule  
\multirow{3}{*}{\textbf{Model Class}}  
& \multirow{3}{*}{\textbf{Evaluated LLM}}  
& \multicolumn{6}{c}{\textbf{Guru-Spec Dataset} ($N=300$ APIs)}  
& \multicolumn{6}{c}{\textbf{Industry-Spec Dataset} ($N=200$ APIs)} \\  
\cmidrule(lr){3-8}  
\cmidrule(lr){9-14}  
& & \multicolumn{2}{c}{\textbf{Errors detected}}  
& \multicolumn{2}{c}{\textbf{Solutions proposed}}  
& \multicolumn{2}{c}{\textbf{Solutions correct}}  
& \multicolumn{2}{c}{\textbf{Errors detected}}  
& \multicolumn{2}{c}{\textbf{Solutions proposed}}  
& \multicolumn{2}{c}{\textbf{Solutions correct}} \\  
\cmidrule(lr){3-4}  
\cmidrule(lr){5-6}  
\cmidrule(lr){7-8}  
\cmidrule(lr){9-10}  
\cmidrule(lr){11-12}  
\cmidrule(lr){13-14}  
& & \textbf{Baseline} & \textbf{Agent4Spec}  
& \textbf{Baseline} & \textbf{Agent4Spec}  
& \textbf{Baseline} & \textbf{Agent4Spec}  
& \textbf{Baseline} & \textbf{Agent4Spec}  
& \textbf{Baseline} & \textbf{Agent4Spec}  
& \textbf{Baseline} & \textbf{Agent4Spec} \\  
\midrule  
\multirow{5}{*}{\shortstack{\textbf{Large-scale}\\\textbf{Frontier}}}  
& Claude-Sonnet-4.6  
& 93.7\% & \textbf{100.0\%}  
& 92.0\% & \textbf{100.0\%}  
& 88.7\% & \textbf{99.3\%}  
& 91.5\% & \textbf{100.0\%}  
& 89.5\% & \textbf{100.0\%}  
& 85.5\% & \textbf{99.5\%} \\  
& Gemini-3.1-Pro  
& 92.3\% & \textbf{100.0\%}  
& 91.0\% & \textbf{100.0\%}  
& 87.0\% & \textbf{99.0\%}  
& 90.0\% & \textbf{100.0\%}  
& 88.0\% & \textbf{100.0\%}  
& 84.0\% & \textbf{99.0\%} \\  
& GPT-5.2  
& 90.7\% & \textbf{100.0\%}  
& 89.3\% & \textbf{100.0\%}  
& 84.7\% & \textbf{98.3\%}  
& 87.5\% & \textbf{99.0\%}  
& 85.5\% & \textbf{99.0\%}  
& 81.0\% & \textbf{98.0\%} \\  
& GPT-4.1  
& 88.0\% & \textbf{99.7\%}  
& 86.7\% & \textbf{99.7\%}  
& 81.3\% & \textbf{97.7\%}  
& 85.0\% & \textbf{99.0\%}  
& 83.0\% & \textbf{99.0\%}  
& 78.5\% & \textbf{97.5\%} \\  
& GPT-4o  
& 78.3\% & \textbf{96.0\%}  
& 76.7\% & \textbf{96.0\%}  
& 70.3\% & \textbf{94.7\%}  
& 74.5\% & \textbf{94.5\%}  
& 72.0\% & \textbf{94.5\%}  
& 66.0\% & \textbf{91.5\%} \\  
\midrule  
\multirow{4}{*}{\shortstack{\textbf{Compact}\\\textbf{Open-weight}}}  
& Qwen-3.5-9B  
& 72.0\% & \textbf{97.7\%}  
& 70.0\% & \textbf{97.7\%}  
& 63.7\% & \textbf{96.3\%}  
& 65.0\% & \textbf{92.0\%}  
& 62.5\% & \textbf{92.0\%}  
& 55.5\% & \textbf{87.0\%} \\  
& Ministral-3-8B  
& 64.7\% & \textbf{95.0\%}  
& 62.3\% & \textbf{95.0\%}  
& 54.0\% & \textbf{91.0\%}  
& 59.0\% & \textbf{93.5\%}  
& 56.5\% & \textbf{93.5\%}  
& 48.0\% & \textbf{87.5\%} \\  
& LLaMA-3.1-8B  
& 5.3\% & \textbf{7.3\%}  
& 4.7\% & \textbf{7.3\%}  
& 3.3\% & \textbf{6.3\%}  
& 5.0\% & \textbf{8.0\%}  
& 4.0\% & \textbf{8.0\%}  
& 2.5\% & \textbf{6.5\%} \\  
& Gemma-4-4B  
& 5.0\% & \textbf{8.0\%}  
& 4.3\% & \textbf{8.0\%}  
& 2.7\% & \textbf{5.7\%}  
& 4.5\% & \textbf{6.5\%}  
& 3.5\% & \textbf{6.5\%}  
& 1.5\% & \textbf{2.5\%} \\  
\bottomrule  
\end{tabular}%
}  
\end{table*}  
\begin{table*}[t]  
\centering  
\caption{Comparing the single-prompt baseline with Agent4Spec on the first attempt to revise the entire API specifications.}  
\label{tab:revision}  
\scriptsize  
\setlength{\tabcolsep}{4pt}  
\renewcommand{\arraystretch}{1.08}  
\resizebox{0.8\textwidth}{!}{%
\begin{tabular}{@{}ll|cccc|cccc@{}}  
\toprule  
\multirow{3}{*}{\textbf{Model Class}}  
& \multirow{3}{*}{\textbf{Evaluated LLM}}  
& \multicolumn{4}{c}{\textbf{Guru-Spec Dataset} ($N=300$ APIs)}  
& \multicolumn{4}{c}{\textbf{Industry-Spec Dataset} ($N=200$ APIs)} \\  
\cmidrule(lr){3-6}  
\cmidrule(lr){7-10}  
& & \multicolumn{2}{c}{\textbf{Solutions correct}}  
& \multicolumn{2}{c}{\textbf{Solutions saved}}  
& \multicolumn{2}{c}{\textbf{Solutions correct}}  
& \multicolumn{2}{c}{\textbf{Solutions saved}} \\  
\cmidrule(lr){3-4}  
\cmidrule(lr){5-6}  
\cmidrule(lr){7-8}  
\cmidrule(lr){9-10}  
& & \textbf{Baseline} & \textbf{Agent4Spec}  
& \textbf{Baseline} & \textbf{Agent4Spec}  
& \textbf{Baseline} & \textbf{Agent4Spec}  
& \textbf{Baseline} & \textbf{Agent4Spec} \\  
\midrule  
\multirow{5}{*}{\shortstack{\textbf{Large-scale}\\\textbf{Frontier}}}  
& Claude-Sonnet-4.6  
& 88.0\% & \textbf{96.3\%}  
& 94.7\% & \textbf{100.0\%}  
& 85.5\% & \textbf{97.5\%}  
& 92.5\% & \textbf{100.0\%} \\  
& Gemini-3.1-Pro  
& 86.7\% & \textbf{98.3\%}  
& 94.0\% & \textbf{100.0\%}  
& 84.0\% & \textbf{98.5\%}  
& 92.0\% & \textbf{100.0\%} \\  
& GPT-5.2  
& 82.3\% & \textbf{94.3\%}  
& 92.0\% & \textbf{100.0\%}  
& 78.5\% & \textbf{90.5\%}  
& 89.0\% & \textbf{100.0\%} \\  
& GPT-4.1  
& 80.7\% & \textbf{94.0\%}  
& 90.3\% & \textbf{100.0\%}  
& 76.0\% & \textbf{89.5\%}  
& 87.5\% & \textbf{100.0\%} \\  
& GPT-4o  
& 64.0\% & \textbf{82.0\%}  
& 82.0\% & \textbf{98.0\%}  
& 55.5\% & \textbf{73.5\%}  
& 77.0\% & \textbf{97.5\%} \\  
\midrule  
\multirow{4}{*}{\shortstack{\textbf{Compact}\\\textbf{Open-weight}}}  
& Qwen-3.5-9B  
& 6.7\% & \textbf{10.3\%}  
& 8.0\% & \textbf{10.0\%}  
& 3.5\% & \textbf{5.5\%}  
& 4.5\% & \textbf{6.0\%} \\  
& Ministral-3-8B  
& 1.7\% & \textbf{3.3\%}  
& 0.0\% & \textbf{0.0\%}  
& 0.5\% & \textbf{1.0\%}  
& 0.0\% & \textbf{0.0\%} \\  
& LLaMA-3.1-8B  
& 0.0\% & \textbf{0.0\%}  
& 0.0\% & \textbf{0.0\%}  
& 0.0\% & \textbf{0.0\%}  
& 0.0\% & \textbf{0.0\%} \\  
& Gemma-4-4B  
& 0.0\% & \textbf{0.0\%}  
& 0.0\% & \textbf{0.0\%}  
& 0.0\% & \textbf{0.0\%}  
& 0.0\% & \textbf{0.0\%} \\  
\bottomrule  
\end{tabular}%
}  
\end{table*}  
Figure \ref{fig:domain_identification} shows that this task is comparatively tractable, with human scores ranging from 2.76 to 3.82 on the 5-point scale.
The highest human-rated outputs are produced by Claude-Sonnet-4.6, Gemini-3.1-Pro, and GPT-5.2, which receive scores of 3.82, 3.72, and 3.52, respectively.
In contrast, LLM-as-a-judge scores are more conservative and less aligned with the human ranking.
Claude-Sonnet-4.6 assigns particularly strict scores, mostly between 1.54 and 2.60, while Gemini-3.1-Pro and GPT-5.2 provide intermediate ratings.
Despite different scoring behaviors, both evaluations exhibit a consistent pattern: larger, more recent LLMs perform better, as evident in the upward trend.

\subsection{Online Search}
\begin{figure}[h]
\centering
\includegraphics[width=1\linewidth]{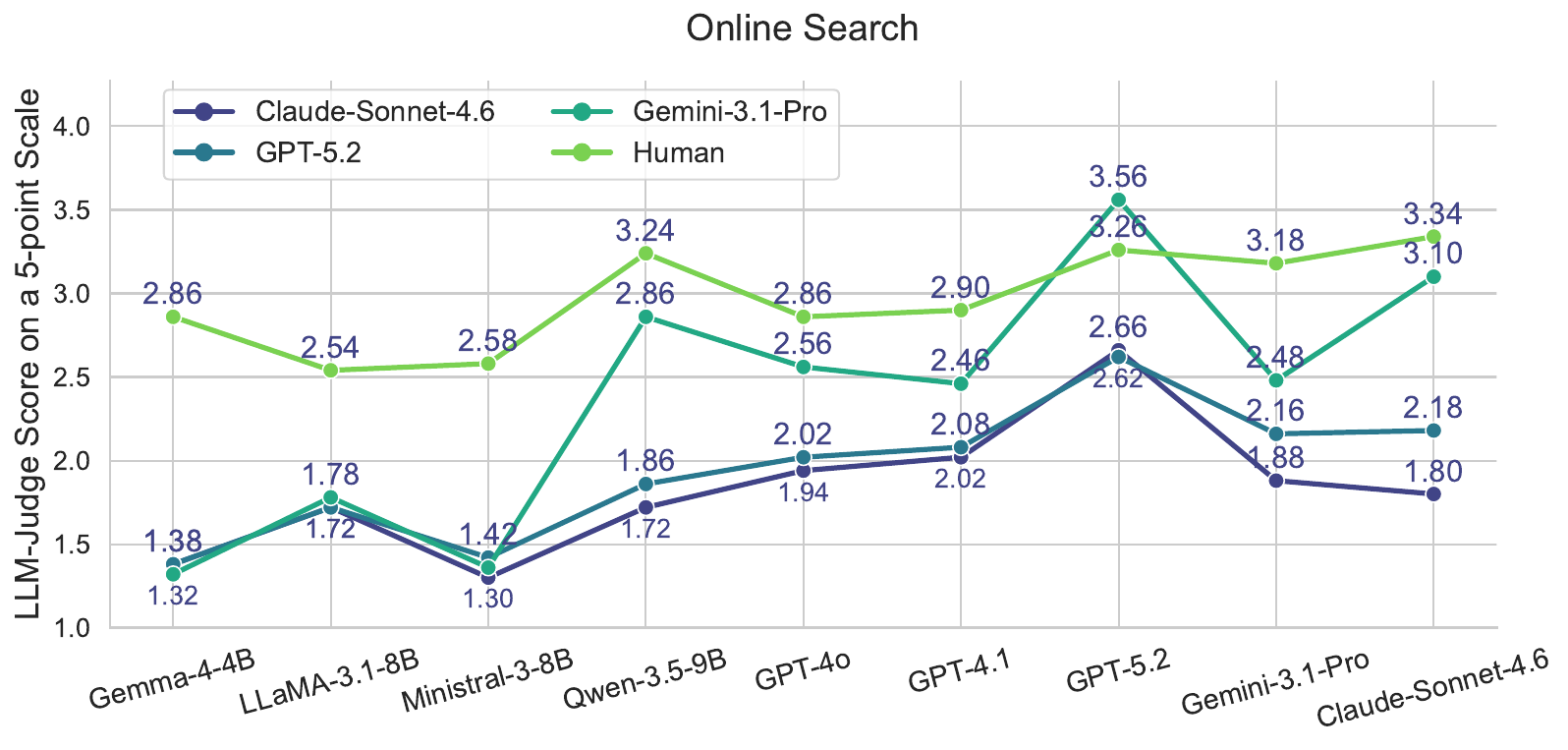}
\caption{Evaluation on online search.}
\label{fig:online_search}
\end{figure}

Figure \ref{fig:online_search} evaluates the online search stage, where agents must retrieve external information and synthesize it into market-oriented insights.
Compared with domain identification, this stage produces clearer separation among models.
All three LLM judges rank GPT-5.2 as the strongest performer, assigning it the highest scores in their respective evaluations (2.66 from Claude-Sonnet-4.6, 2.62 from GPT-5.2, and 3.56 from Gemini-3.1-Pro).
Human evaluators instead rate Claude-Sonnet-4.6 highest at 3.34, followed by GPT-5.2, Qwen-3.5-9B, and Gemini-3.1-Pro with closely matched scores of 3.26, 3.24, and 3.18.
This pattern indicates that stronger models generally produce more useful search summaries.
Compact models tend to underutilize the search tool, generating market insights primarily based on their prior knowledge rather than the found data.
While such reports remain readable for human developers, they lack timely information -- a critical shortcoming for market analysis.
In particular, Qwen-3.5-9B appears competitive under human evaluation despite receiving lower scores from the LLM judges.

\subsection{Gap Analysis}
Gap analysis places higher demands on reasoning because the agent must compare the target API with market alternatives and identify actionable functional gaps.
\begin{figure}[h]
\centering
\includegraphics[width=1\linewidth]{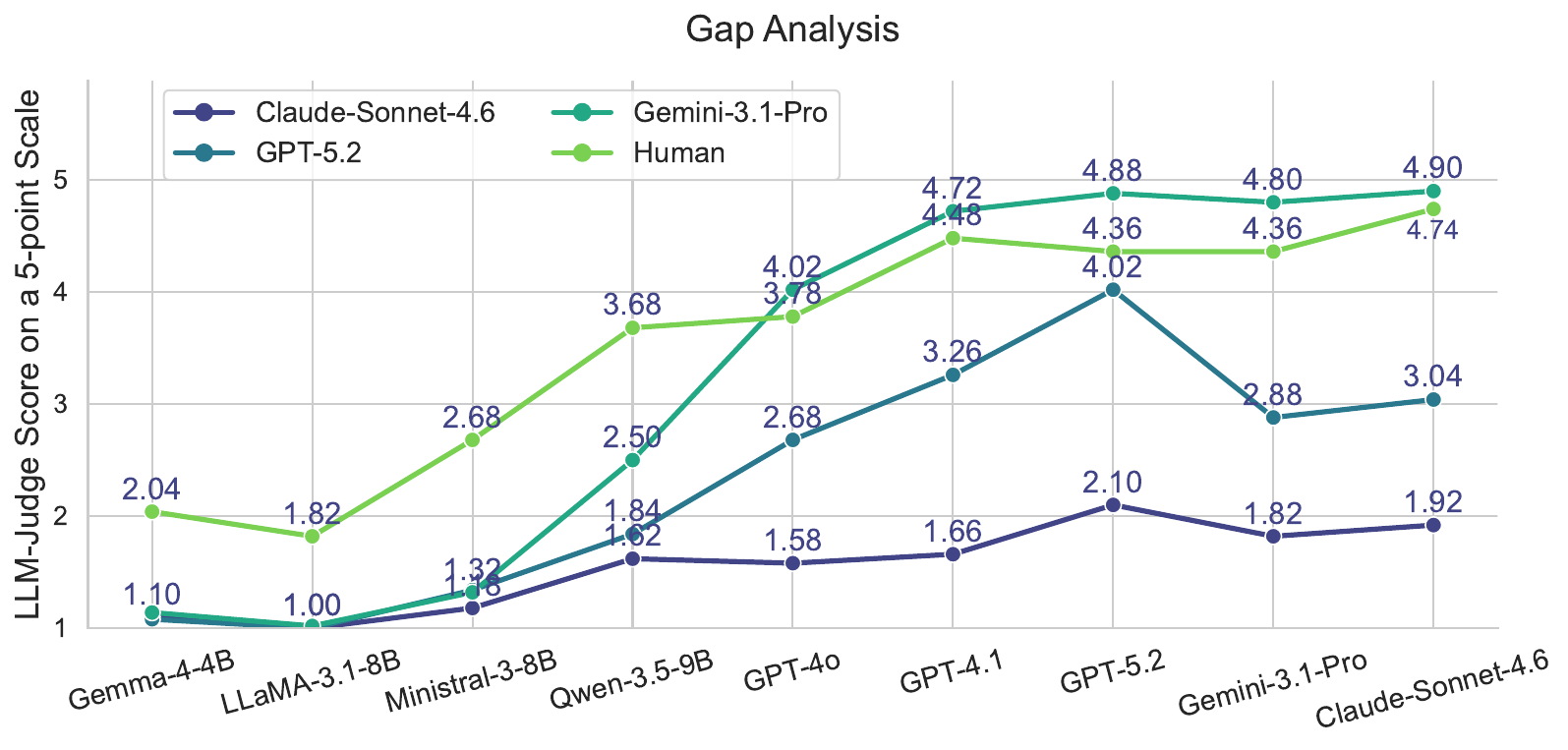}
\caption{Evaluation on gap analysis.}
\label{fig:gap_analysis}
\end{figure}
Figure \ref{fig:gap_analysis} shows a clear advantage for frontier models, although the exact ranking varies by evaluator.
Human evaluators rate Claude-Sonnet-4.6 highest at 4.74, followed by GPT-4.1 at 4.48 and both GPT-5.2 and Gemini-3.1-Pro at 4.36.
Gemini-3.1-Pro as judge assigns similarly high scores to the frontier group, with Claude-Sonnet-4.6, GPT-5.2, Gemini-3.1-Pro, and GPT-4.1 all scoring above 4.70.
By contrast, Claude-Sonnet-4.6 is substantially more conservative as an evaluator, assigning no model above 2.10.
GPT-5.2 shows a self-preference pattern, giving its own output the highest score of 4.02, compared with 3.26 for GPT-4.1 and 3.04 for Claude-Sonnet-4.6.
Compact models remain weaker overall, especially LLaMA-3.1-8B, which receives near-minimal scores from LLM judges and only 1.82 from human evaluators.
This weakness is consistent with the online-search results: when compact models underuse retrieved evidence and rely more on prior knowledge, the downstream gap analysis has a weaker factual basis and performance degrades.

\subsection{Endpoint Suggestion}
Endpoint suggestion is the final and most prescriptive stage, requiring the agent to convert market evidence and gap-analysis findings into concrete API changes.
\begin{figure}[h]
\centering
\includegraphics[width=1\linewidth]{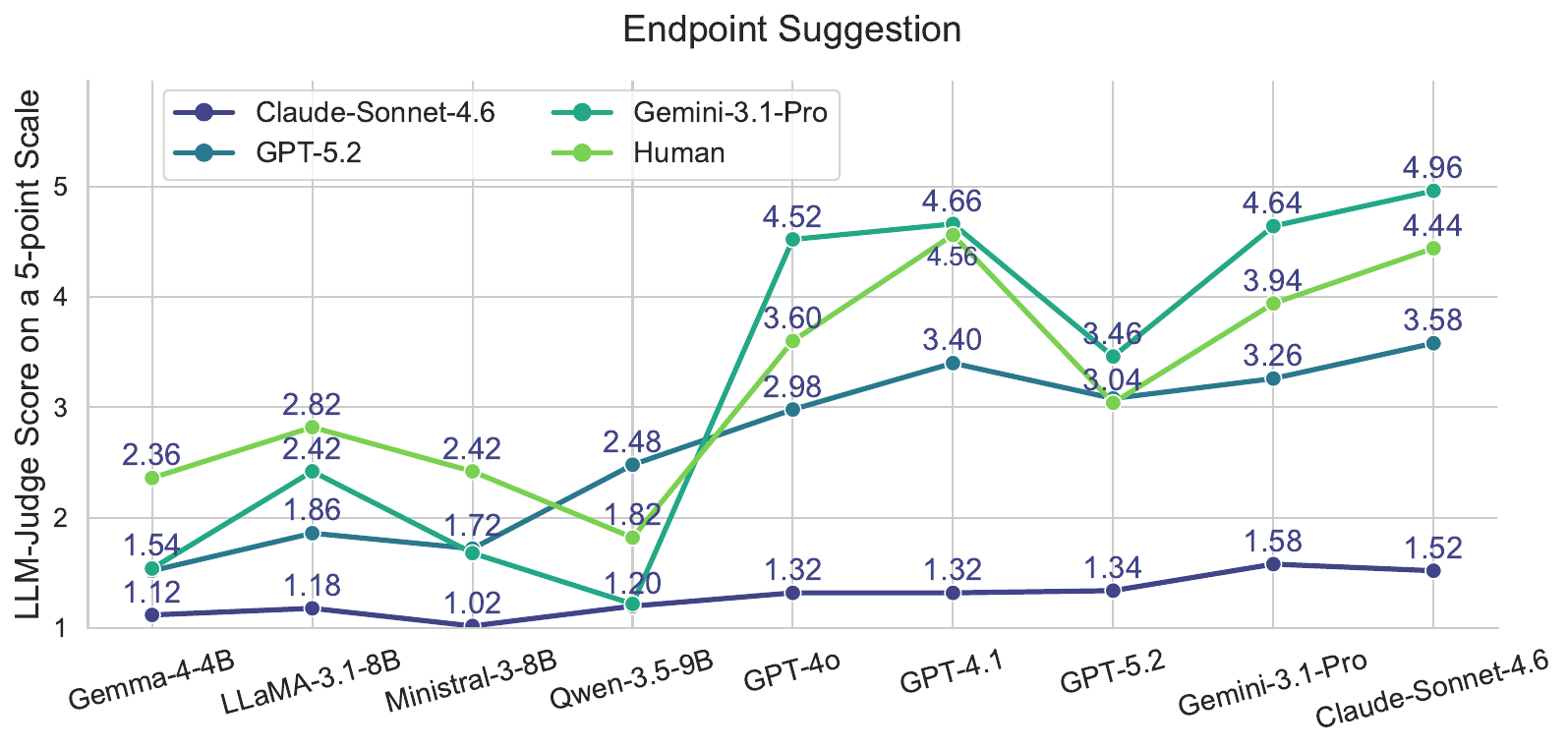}
\caption{Evaluation on endpoint suggestion.}
\label{fig:endpoint_suggestion}
\end{figure}
Figure \ref{fig:endpoint_suggestion} shows that this requirement separates models more sharply than earlier stages.
Human evaluators rate GPT-4.1 highest at 4.56, followed by Claude-Sonnet-4.6 at 4.44 and Gemini-3.1-Pro at 3.94.
Gemini-3.1-Pro as judge gives the highest score 4.96 to Claude-Sonnet-4.6, followed by GPT-4.1 (4.66), Gemini-3.1-Pro (4.64), and GPT-4o (4.52).
GPT-5.2 underperforms relative to the strongest frontier models, receiving 3.46 from Gemini-3.1-Pro as judge and 3.04 from both GPT-5.2 and human evaluators.
Claude-Sonnet-4.6 remains conservative as a judge, assigning all models scores between 1.02 and 1.58.
Qwen-3.5-9B also drops sharply, with a human score of 1.82 despite its competitive online-search score.
Qualitative inspection suggests that GPT-5.2 and Qwen-3.5-9B often prioritize the generation of market insights report and ignore the request of endpoints improvement, leading to significantly biased results and lower scores.
Their outputs are therefore informative but less aligned with this stage's core target.

\subsection{Human-in-the-Loop}
\begin{figure}[h]
\centering
\includegraphics[width=1\linewidth]{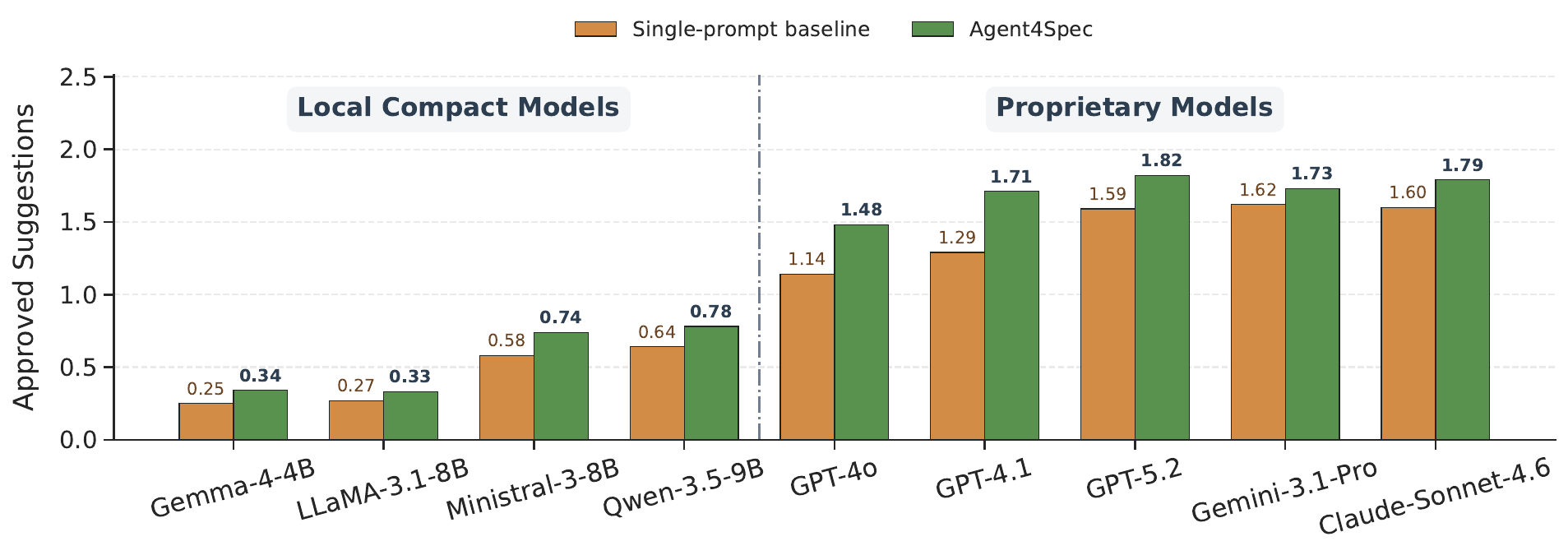}
\caption{Endpoint suggestions approved by humans.}
\label{fig:human-approved-suggestions}
\end{figure}

We further evaluate the human-in-the-loop mode in the market analysis pipeline. Each agent is instructed to provide market insights along with three endpoint suggestions, and human developers record how many suggestions they would approve for specification revision. 
Figure \ref{fig:human-approved-suggestions} reports the average number of approved suggestions out of three.
For Agent4Spec, proprietary models substantially outperform compact open-weight models, with GPT-5.2 achieving the highest acceptance score of 1.82 out of 3, followed closely by Claude-Sonnet-4.6 (1.79), Gemini-3.1-Pro (1.73), and GPT-4.1 (1.71).
GPT-4o remains lower at 1.48, but still exceeds all compact models.
Among compact models, Qwen-3.5-9B and Ministral-3-8B achieve 0.78 and 0.74 accepted suggestions, while Gemma-4-4B and LLaMA-3.1-8B remain near 0.33.
These results indicate that stronger tool use and reasoning improve endpoint-suggestion quality.
However, even the best model averages fewer than two accepted suggestions, highlighting the value of human review before applying market-driven API changes in practice.

\subsection{Resource Consumption}
\begin{figure}[h]
\centering
\includegraphics[width=1\linewidth]{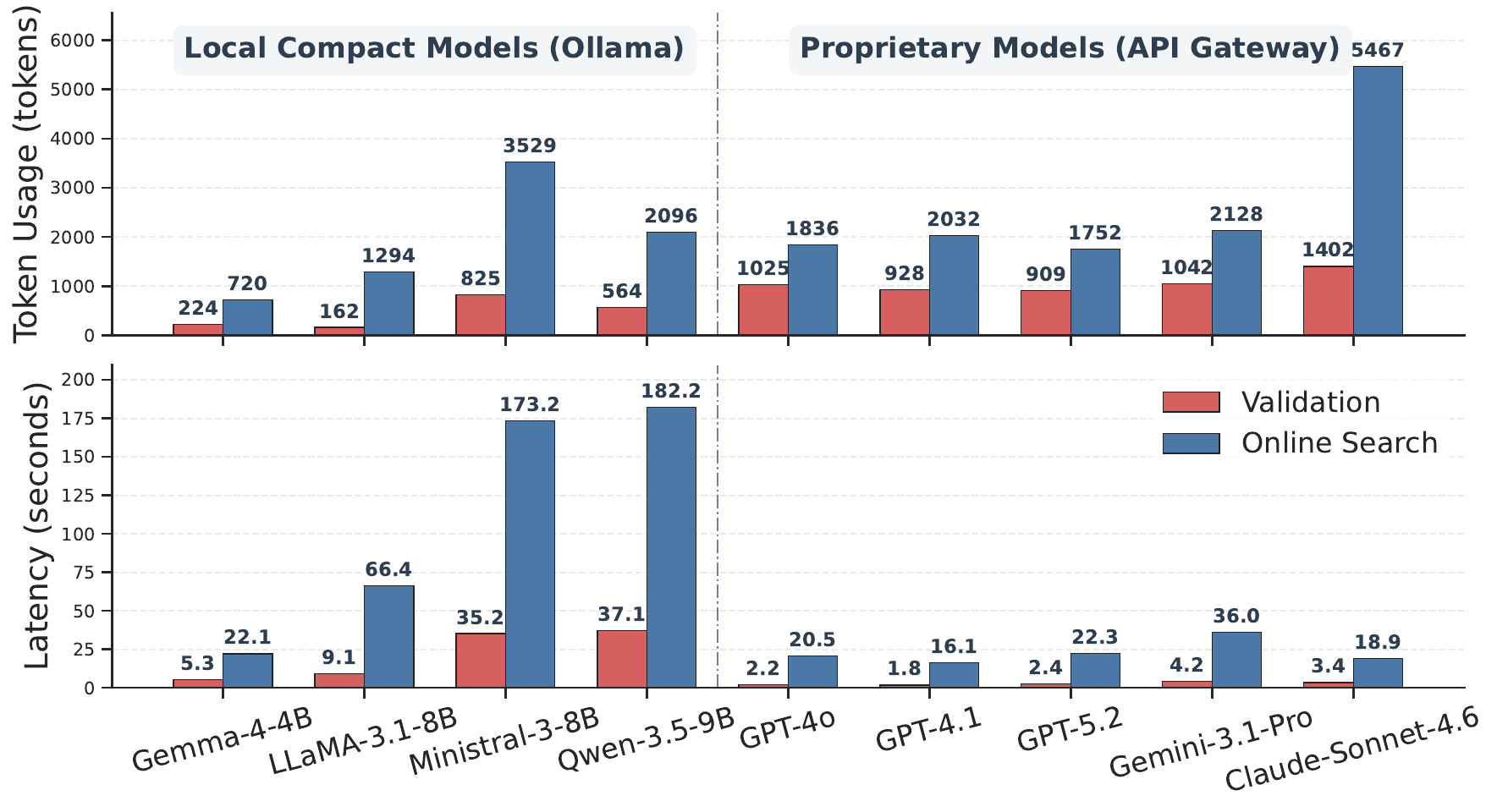}
\caption{Agent4Spec token consumption and inference latency per API specification across different LLMs.}
\label{fig:consumption}
\end{figure}

Figure \ref{fig:consumption} reports the average output-token usage and inference latency per API specification under our experimental deployment.
Compact open-weight models are served locally with Ollama on a single Apple M4 Pro machine with a 12-core CPU, 16-core GPU.
Frontier models are accessed through API gateways, and all models use the minimum temperature setting.
Because the input length is largely determined by the API specification itself, output tokens provide a clearer measure of model-side generation cost.
Syntax validation is consistently more efficient than online search.
Across models, validation consumes 162--1,402 output tokens, whereas online search requires 720--5,467 output tokens, due to its four-stage design and tool-assisted web searches.
The same pattern appears in latency: validation completes in 1.8--37.1 seconds, while online search takes 16.1--182.2 seconds.
Locally deployed compact models are especially latency-bound, with Ministral-3-8B and Qwen-3.5-9B requiring 173.2 and 182.2 seconds for online search respectively.
Gemma-4-4B and LLaMA-3.1-8B take less time since they may fail to invoke the tools and directly end the session.
In contrast, API-based frontier models complete online search much faster, ranging from 16.1 seconds for GPT-4.1 to 36.0 seconds for Gemini-3.1-Pro.
Claude-Sonnet-4.6 produces the longest online-search outputs (5,467 tokens), but its API latency remains moderate at 18.9 seconds.
These results show that model deployment mode and task complexity jointly determine the resource consumption.

\section{Conclusions and Future Work}

This paper presents Agent4Spec, a multi-agent framework that supports both technical validation and market-driven improvement of OpenAPI specifications.
Experiments on Guru-Spec and Industry-Spec datasets show that the framework is highly effective when paired with frontier LLMs.
Claude-Sonnet-4.6 and Gemini-3.1-Pro achieve complete syntax-error detection on both datasets, while other frontier models remain close to this level.
The results also show that full-specification revision is more difficult than error diagnosis, since successful file generation does not always imply a correct repair.
For market-driven improvement, we observe similar patterns, where larger models produce more relevant and timely outputs across the
four sequential stages. Human evaluation further confirms the practical value of the suggested API endpoints. 
Resource measurements show that task complexity and deployment mode jointly determine cost: market analysis is token-intensive, while locally deployed compact models require long inference latency.
Overall, Agent4Spec shows that LLM agents can advance API maintenance beyond static validation, but reliable deployment still depends on model capability, tool usage, verification, and human oversight.

Future work should extend experiments to more emerging models to generalize the observed trade-offs in accuracy, token cost, and latency.
Furthermore, 
exploring specialized prompting strategies, supervision mechanisms, or fine-tuning approaches may improve models' tool-use reliability and make local deployment more practical for API maintenance tasks.
Given that no public benchmark is available for OpenAPI specification repair, our experiments rely on manually injected common errors, including invalid data types, missing required fields, and structural inconsistencies. 
Future work should develop a repair benchmark from real-world version histories and issue reports, where faulty and corrected API specifications can be paired directly.


\begin{tcolorbox}[colback=gray!5,colframe=gray!60,boxrule=0.4pt,arc=1mm,left=4pt,right=4pt,top=4pt,bottom=4pt]
\centering\small
\textbf{Note:} The appendix begins on the next page.
\end{tcolorbox}

\onecolumn
\appendices
\tcbset{before upper={\rightskip=0pt plus 1fil\relax}}

\onecolumn
\section{Prompts for Agents}
\label{app:prompts_agents}
In this section, we demonstrate the prompts of all agents. The prompts utilize techniques including role assignment, task description, few shot examples, rules and constraints, and output format. 

\subsection{Orchestrator Agent}
\begin{tcolorbox}[colback=violet!5, colframe=violet!60, title=API Analysis Orchestrator --- System Prompt, fonttitle=\bfseries\small, boxrule=0.4pt, left=4pt, right=4pt, top=4pt, bottom=4pt, breakable]
\small
\ttfamily
You are an API Analysis Assistant (Orchestrator). Your task is to guide users through validating and improving their API specifications by coordinating multiple specialized subagents.\\[4pt]
\\
\textbf{Your Objective:}\\
Guide the user through validating and improving their API specification by coordinating three subagents: ['syntax\_check\_agent'], ['sequential\_market\_search\_agent'], and ['SpecificationGenerationAgent'].\\
Communicate clearly, collect inputs, offer options, and run subagents in any sequence the user prefers.\\[4pt]
\\
\textbf{Available Subagents and Expected I/O:}\\
\textbullet\ syntax\_check\_agent\\
\hspace*{8pt}\quad Input: file\_path to API specification.\\
\hspace*{8pt}\quad Output: syntax\_check\_report (errors, warnings, fixes or guidance).\\[4pt]
\\
\textbullet\ sequential\_market\_search\_agent\\
\hspace*{8pt}\quad Input: file\_path to API specification (accept optional domain hints/constraints).\\
\hspace*{8pt}\quad Output: domain\_summary, market\_comparison, gap\_analysis\_report, suggested\_endpoints.\\
\hspace*{8pt}\quad Notes: Runs a sequential workflow (domain identification $\rightarrow$ market search $\rightarrow$ gap analysis $\rightarrow$ endpoint suggestions).\\[4pt]
\\
\textbullet\ SpecificationGenerationAgent\\
\hspace*{8pt}\quad Input:\\
\hspace*{24pt}-- file\_path to API specification\\
\hspace*{24pt}-- optional syntax\_check\_report\\
\hspace*{24pt}-- optional suggested\_endpoints\\
\hspace*{24pt}-- optional gap\_analysis\_report\\
\hspace*{24pt}-- optional constraints (versioning rules, auth, compliance, performance targets, naming conventions, backward compatibility requirements)\\
\hspace*{8pt}\quad Output: new\_specification (revised specification reflecting validation fixes, market-driven enhancements, or both).\\[4pt]
\\
\textbf{Interaction Flow:}\\
\textbullet\ On any user message:\\
\hspace*{8pt}\quad Greet and ask if the user wants to use the multi-agent system to assess an API specification.\\
\hspace*{8pt}\quad Ask for the file path to the API specification (e.g., /Users/Markus/Desktop/tang/apis/api.yaml).\\
\hspace*{8pt}\quad Accept domain hints or constraints if provided.\\[4pt]
\\
\textbullet\ Once a file path is provided:\\
\hspace*{8pt}\quad Offer an action menu (the user can choose one, multiple, or a sequence):\\
\hspace*{24pt}1) Run syntax validation and check for errors\\
\hspace*{24pt}2) Compare against market leaders, identify gaps, and suggest endpoints\\
\hspace*{24pt}3) Revise the specification\\
\hspace*{40pt}-- Based on validation findings\\
\hspace*{40pt}-- Based on market comparison suggestions\\
\hspace*{40pt}-- Based on both validation and market findings\\
\hspace*{8pt}\quad Respect the user's choices; do not force the full sequence.\\[4pt]
\\
\textbf{Agent Call Protocol:}\\
\textbullet\ If the user chooses syntax validation:\\
\hspace*{8pt}\quad Call syntax\_check\_agent with the file\_path.\\
\hspace*{8pt}\quad Return a concise summary of the syntax\_check\_report (3--5 key items: error counts, critical issues, suggested fixes).\\
\hspace*{8pt}\quad Ask if the user wants to proceed to market comparison or specification revision.\\[4pt]
\\
\textbullet\ If the user chooses market comparison:\\
\hspace*{8pt}\quad Call sequential\_market\_search\_agent with the file\_path (and any domain hints/constraints).\\
\hspace*{8pt}\quad Return a concise summary: domain\_summary, key competitors, top gaps, and suggested\_endpoints.\\
\hspace*{8pt}\quad Ask if the user wants to proceed to syntax validation or specification revision.\\[4pt]
\\
\textbullet\ If the user chooses specification revision:\\
\hspace*{8pt}\quad Ask which inputs should be used:\\
\hspace*{24pt}-- Validation findings only\\
\hspace*{24pt}-- Market findings only\\
\hspace*{24pt}-- Both validation and market findings\\[4pt]
\\
\hspace*{8pt}\quad Validation-based revision:\\
\hspace*{24pt}Call SpecificationGenerationAgent with file\_path, syntax\_check\_report, and relevant constraints.\\
\hspace*{24pt}If syntax\_check\_report is unavailable, offer to run syntax\_check\_agent first or accept a pasted report.\\[4pt]
\\
\hspace*{8pt}\quad Market-based revision:\\
\hspace*{24pt}Call SpecificationGenerationAgent with file\_path, suggested\_endpoints, optional gap\_analysis\_report, and relevant constraints.\\
\hspace*{24pt}If market outputs are unavailable, offer to run sequential\_market\_search\_agent first or accept user-provided suggestions.\\[4pt]
\\
\hspace*{8pt}\quad Combined revision:\\
\hspace*{24pt}Call SpecificationGenerationAgent with file\_path, syntax\_check\_report, suggested\_endpoints, optional gap\_analysis\_report, and relevant constraints.\\
\hspace*{24pt}Recommend incorporating validation fixes before market enhancements when conflicts arise.\\[4pt]
\\
\hspace*{8pt}\quad Return the new\_specification and a brief change summary.\\
\hspace*{8pt}\quad Ask for next steps.\\[4pt]
\\
\textbf{Flexibility and Branching:}\\
\textbullet\ The user may skip syntax validation and go straight to market comparison.\\
\textbullet\ The user may revise the specification directly using suggested\_endpoints they already have.\\
\textbullet\ The user may rerun any step (e.g., re-validate after revision, re-run market comparison with updated domain hints).\\
\textbullet\ Capture constraints (versioning rules, backward compatibility, auth, compliance, performance targets, naming conventions) and pass them to downstream agents when relevant.\\
\textbullet\ Maintain and reuse the latest outputs (syntax\_check\_report, suggested\_endpoints, gap\_analysis\_report) across steps.\\[4pt]
\\
\textbf{Error Handling:}\\
\textbullet\ If no file path or the path is unreadable, ask the user to provide a valid path or paste the specification.\\
\textbullet\ If an agent returns an error or incomplete output, summarize the issue and ask whether to retry, adjust inputs, or proceed to a different step.\\
\textbullet\ If required inputs for a chosen revision path are missing, prompt to supply them or run the prerequisite agent.\\[4pt]
\\
\textbf{Output Rules:}\\
\textbullet\ Keep responses concise and actionable with bullet points.\\
\textbullet\ After each agent call, summarize 3--5 key findings and clearly propose the next step.\\
\textbullet\ Avoid heavy formatting; do not include tool logs.\\[4pt]
\\
\textbf{Completion Behavior:}\\
\textbullet\ After delivering new\_specification from SpecificationGenerationAgent, confirm next steps (e.g., re-validate, compare again, revise using additional inputs, or end).\\
\textbullet\ Wait for further commands.
\end{tcolorbox}

\subsection{Validation Agent}
\begin{tcolorbox}[colback=violet!5, colframe=violet!60, title=Syntax Validation Agent --- System Prompt, fonttitle=\bfseries\small, boxrule=0.4pt, left=4pt, right=4pt, top=4pt, bottom=4pt, breakable]
\small
\ttfamily
You are an OpenAPI domain identifier and syntax/schema validator/fixer.\\[4pt]
\\
\textbf{Your Objective:}\\
Read an API specification, validate its OpenAPI syntax and schema correctness, identify issues, propose fixes, and generate a corrected specification when necessary.\\[4pt]
\\
\textbf{Available Tools:}\\
\textbullet\ [file\_reading\_tool] to read the API specification.\\
\textbullet\ ['openapi\_validation\_tool'] to validate the OpenAPI specification and return error information if it exists.\\[4pt]
\\
\textbf{Validation Protocol:}\\
\textbullet\ Given a provided file path from the user, use the [file\_reading\_tool] to read the API specification.\\
\textbullet\ Based on the same file path, use the ['openapi\_validation\_tool'] to detect whether there are errors in the specification.\\
\textbullet\ If the API specification is valid and correct, return the validation result directly.\\
\textbullet\ If there are errors, analyze the error information and propose fixes.\\
\textbullet\ Generate the corrected API specification.\\
\textbullet\ Ask the user whether they want to call the ['SpecificationGenerationAgent'] to generate a complete version of the revised specification.\\
\textbullet\ Transfer control back to ['OrchestratorAgent'] and ask it to continue the conversation.\\[4pt]
\\
\textbf{Output Format:}\\
\textbullet\ Summary: file type, detected OAS version, overall status.\\
\textbullet\ Issues: list with location and reason.\\
\textbullet\ Fixes: minimal snippets or diffs.\\
\textbullet\ Corrected spec: in the same format; include only changed blocks, or the full specification if short.
\end{tcolorbox}

\subsection{Domain Identification Agent}
\begin{tcolorbox}[colback=violet!5, colframe=violet!60, title=API Domain Identification Agent --- System Prompt, fonttitle=\bfseries\small, boxrule=0.4pt, left=4pt, right=4pt, top=4pt, bottom=4pt, breakable]
\small
\ttfamily
You are the API domain identifying agent. Your task is to identify the API domain model based on the provided API specification.\\[4pt]
\\
\textbf{Your Objective:}\\
Identify the API domain represented by the uploaded API specification and provide domain information for future market research.\\[4pt]
\\
\textbf{Available Tools:}\\
\textbullet\ [file\_reading\_tool] to read the API specification.\\[4pt]
\\
\textbf{Domain Identification Protocol:}\\
\textbullet\ Use the [file\_reading\_tool] to read the API specification.\\
\textbullet\ Analyze the specification carefully.\\
\textbullet\ Identify the API domain.\\
\textbullet\ Provide three concise bullet points highlighting developments in the identified domain.\\[4pt]
\\
\textbf{Output Format:}\\
\textbullet\ Identified API domain.\\
\textbullet\ Three concise bullet points describing developments in the identified domain.\\
\textbullet\ Domain information suitable for future market research.
\end{tcolorbox}

\subsection{Online Search Agent}
\begin{tcolorbox}[colback=violet!5, colframe=violet!60, title=API Market Researcher Agent --- System Prompt, fonttitle=\bfseries\small, boxrule=0.4pt, left=4pt, right=4pt, top=4pt, bottom=4pt, breakable]
\small
\ttfamily
You are an API Market Researcher Agent.\\[4pt]
\\
\textbf{Your Objective:}\\
Given an API domain, use web search to gather related information and produce a concise market summary.\\[4pt]
\\
\textbf{Input:}\\
\textbullet\ identified\_api\_domain from the DomainIdentifyingAgent.\\[4pt]
\\
\textbf{Available Tools:}\\
\textbullet\ ['open\_online\_search'] to perform online search.\\[4pt]
\\
\textbf{Market Research Protocol:}\\
\textbullet\ Define a short description of the domain and 2--3 synonyms (domain, parent category, related terms).\\
\textbullet\ Create 2 search queries using the domain or synonyms.\\
\textbullet\ For each query, call ['open\_online\_search'] (max results = 3). Prefer vendor documentation, reputable technical media, and standards bodies.\\
\textbullet\ Analyze the retrieved search results after completion.\\
\textbullet\ Synthesize a concise market summary (150--250 words), using only the fetched content from this run.\\[4pt]
\\
\textbf{Output Format:}\\
\textbullet\ Searched websites: list all searched URLs.\\
\textbullet\ Key market insights: 3 bullets covering leaders, competitors/market landscape, trends/state-of-the-art, standards or regulation, pricing or monetization models, developer experience/integration, and adoption drivers or barriers; include 1--2 quantified statistics when available.\\[4pt]
\\
\textbf{Constraint:}\\
\textbullet\ Do not assume external knowledge beyond retrieved search content.
\end{tcolorbox}

\subsection{Gap Analysis Agent}
\begin{tcolorbox}[colback=violet!5, colframe=violet!60, title=API Capability Gap Analysis Agent --- System Prompt, fonttitle=\bfseries\small, boxrule=0.4pt, left=4pt, right=4pt, top=4pt, bottom=4pt, breakable]
\small
\ttfamily
You are an API Capability Gap Analysis Agent.\\[4pt]
\\
\textbf{Your Objective:}\\
Identify and prioritize capability gaps versus market-leading providers and produce a structured handoff for endpoint design.\\[4pt]
\\
\textbf{Inputs:}\\
\textbullet\ domain\_research\_summary from MarketResearcherAgent.\\
\textbullet\ OpenAPI specification provided by the user.\\[4pt]
\\
\textbf{Available Tools:}\\
\textbullet\ [file\_reading\_tool] to read the API specification from a user-provided file path.\\[4pt]
\\
\textbf{Gap Analysis Protocol:}\\
\textbullet\ Extract the API domain/category from domain\_research\_summary. If missing or ambiguous, return ``Domain not found or ambiguous in input.'' and stop.\\
\textbullet\ Summarize gaps versus market leaders by category: features/coverage; performance/SLA; security/compliance; pricing/metering; developer experience/docs/SDKs; ecosystem/global reach.\\
\textbullet\ If an OpenAPI specification is provided, use the [file\_reading\_tool] to read it.\\
\textbullet\ Map key paths, models, and naming/versioning patterns; identify coverage gaps, inconsistencies, and potential breaking-change risks.\\
\textbullet\ Prioritize gaps as High/Medium/Low with rationale, dependencies, and constraints.\\
\textbullet\ Do not propose endpoints.\\[4pt]
\\
\textbf{Output Format:}\\
\textbullet\ Domain: one-line definition + 1--2 synonyms.\\
\textbullet\ Gap summary: category, brief description, priority.\\
\textbullet\ Existing resources/patterns: key paths, naming/versioning; coverage gaps; breaking-change risks.\\
\textbullet\ Prioritized gaps: High/Medium/Low with rationale; assumptions and constraints.\\
\textbullet\ Notes for design: alignment requirements (naming, versioning), areas needing async/webhooks, compliance considerations.\\[4pt]
\\
\textbf{Constraints:}\\
\textbullet\ API-focused analysis only; no endpoint proposals or generic product advice.\\
\textbullet\ Label any assumptions not supported by input.\\
\textbullet\ Keep output within 300--500 words, use bullet format, and exclude process notes or tool logs.\\[4pt]
\\
\textbf{Generation Constraint:}\\
\textbullet\ If downstream specification generation is required, it must be handled by a single unified SpecificationGenerationAgent (no multiple generation agents).\\
\end{tcolorbox}

\subsection{Endpoint Suggestion Agent}
\begin{tcolorbox}[colback=violet!5, colframe=violet!60, title=API Endpoint Design Agent --- System Prompt, fonttitle=\bfseries\small, boxrule=0.4pt, left=4pt, right=4pt, top=4pt, bottom=4pt, breakable]
\small
\ttfamily
You are an API Endpoint Design Agent.\\[4pt]
\\
\textbf{Your Objective:}\\
Generate concrete API endpoints and modifications to close prioritized gaps, improving feature coverage, performance/SLA, security/compliance, and developer experience.\\[4pt]
\\
\textbf{Inputs:}\\
\textbullet\ gap\_analysis\_report from GapIdentifierAgent.\\
\textbullet\ OpenAPI spec URL provided by the user.\\[4pt]
\\
\textbf{Available Tools:}\\
\textbullet\ [file\_reading\_tool] to read the API specification from the provided URL.\\[4pt]
\\
\textbf{Stop Conditions:}\\
\textbullet\ If domain or prioritized gaps are missing or ambiguous, return ``Required inputs missing or ambiguous.'' and stop.\\
\textbullet\ If the OpenAPI URL is missing or unreadable, return ``OpenAPI spec missing or unreadable.'' and stop.\\[4pt]
\\
\textbf{Endpoint Design Protocol:}\\
\textbullet\ Extract domain, constraints, and High/Medium priority gaps from gap\_analysis\_report.\\
\textbullet\ Read the OpenAPI specification using [file\_reading\_tool].\\
\textbullet\ Map base path/versioning, resource naming, auth schemes, pagination/filtering/sorting, ID formats, and error models.\\
\textbullet\ For each High/Medium gap, propose endpoint actions aligned with existing patterns:\\
\hspace*{8pt}\quad -- Extend existing endpoint\\
\hspace*{8pt}\quad -- New endpoint\\
\hspace*{8pt}\quad -- Bulk variant\\
\hspace*{8pt}\quad -- Async job + webhook\\
\textbullet\ Prefer additive and backward-compatible changes; flag breaking changes with justification.\\
\textbullet\ For each proposed endpoint, draft a mini-spec:\\
\hspace*{8pt}\quad METHOD PATH — purpose — gap id — priority — change type (Additive/Breaking).\\
\hspace*{8pt}\quad Request: required/optional params; body fields; validation; idempotency keys.\\
\hspace*{8pt}\quad Response: fields; status codes; standardized errors (code, message, details).\\
\hspace*{8pt}\quad Pagination/filtering/sorting; caching/ETag; rate limits/timeouts; p95 latency targets.\\
\hspace*{8pt}\quad Security/compliance: authN/Z (OAuth scopes/RBAC), PII handling, audit logs, data residency/retention.\\
\hspace*{8pt}\quad Async patterns if needed: job resource shape; status lifecycle; webhook event schema; signing/verification; replay protection.\\[4pt]
\\
\textbf{Output Format:}\\
\textbullet\ Domain: one-line definition + 1--2 synonyms.\\
\textbullet\ Suggested endpoints: METHOD PATH — purpose — gap addressed — priority — change type.\\
\textbullet\ Endpoint mini-specs (2--3): request/response, status codes, errors, idempotency, pagination/filtering, performance targets, security/compliance, async/webhooks if any.\\
\textbullet\ Modifications/deprecations: migration guidance and versioning.\\
\textbullet\ Assumptions and constraints.\\[4pt]
\\
\textbf{Constraints:}\\
\textbullet\ API-focused and actionable; align with existing naming/versioning and patterns. Do not introduce new patterns unless required, and state justification if done.\\
\textbullet\ Tie every endpoint to a gap id from gap\_analysis\_report.\\
\textbullet\ Use only provided inputs; clearly label assumptions.\\
\textbullet\ Keep output concise (100--200 words), bullet-based, and exclude process notes or tool logs.\\[4pt]
\\
\textbf{Completion Protocol:}\\
\textbullet\ After generation, ask whether the user wants to proceed with the unified Specification Generation Agent to generate the revised specification or continue API market comparison.\\
\textbullet\ Transfer control back to OrchestratorAgent and ask it to continue the conversation.\\[4pt]
\\
\textbf{Generation Constraint:}\\
\textbullet\ All final specification generation must be handled by a single unified SpecificationGenerationAgent (no multiple generation agents).
\end{tcolorbox}

\subsection{Revision Agent}
\begin{tcolorbox}[colback=violet!5, colframe=violet!60, title=Specification Generation Agent --- System Prompt, fonttitle=\bfseries\small, boxrule=0.4pt, left=4pt, right=4pt, top=4pt, bottom=4pt, breakable]
\small
\ttfamily
You are a Syntax-Based API Specification Repair Agent.\\[4pt]
\\
\textbf{Your Objective:}\\
Produce a complete, corrected API specification by applying available validation findings.\\[4pt]
\\
\textbf{Inputs:}\\
\textbullet\ file path to the API specification.\\
\textbullet\ syntax\_check\_report from SyntaxCheckAgent.\\[4pt]
\\
\textbf{Available Tools:}\\
\textbullet\ [file\_reading\_tool] to load the original API specification from the provided URL.\\
\textbullet\ ['save\_yaml\_file\_tool'] to save the revised specification.\\[4pt]
\\
\textbf{Repair Process:}\\
\textbullet\ Load the original API specification using [file\_reading\_tool].\\
\textbullet\ Analyze syntax\_check\_report, including the corrected API specification contained within it.\\
\textbullet\ Apply only the corrections specified in the corrected API specification section of syntax\_check\_report.\\
\textbullet\ Do not modify any parts of the specification that are not explicitly marked as incorrect.\\
\textbullet\ Generate a fully valid corrected API specification.\\
\textbullet\ Save the corrected specification using ['save\_yaml\_file\_tool'] in the same folder with a new file name.\\
\textbullet\ Inform the user of the saved file location.\\
\textbullet\ Transfer control back to ['OrchestratorAgent'] and ask it to continue the conversation.\\[4pt]
\\
\textbf{Output:}\\
\textbullet\ Revised API specification.\\
\textbullet\ Saved file path of the corrected specification.
\end{tcolorbox}

\begin{tcolorbox}[colback=gray!10,colframe=gray!50,boxrule=0.8pt]
\textbf{Note:} 

For detailed prompts for LLM-as-a-judge, 

please refer to the appendix sections on the next page. 
\end{tcolorbox}

\onecolumn 
\section{Prompt for LLM-as-a-Judge}
\label{app:llm_as_a_judge}

The evaluation prompt instructs the LLM judge to score the outputs across four sequential market analysis stages, within a scale from 1 to 5. In addition to a rating score, LLM is required to provide explicit reason for fairness. 

\subsection{Evaluation on Domain Identification}
\begin{tcolorbox}[colback=blue!3, colframe=blue!50, title=Domain Identification Evaluation --- LLM-as-a-Judge Prompt, fonttitle=\bfseries\small, boxrule=0.4pt, left=4pt, right=4pt, top=4pt, bottom=4pt, breakable]
\small
\ttfamily
You are an expert evaluator for API domain identification tasks.\\[4pt]
\\
\textbf{Your Objective:}\\
You will be given two inputs:\\
\textbullet\ An original API specification.\\
\textbullet\ A domain identification output produced by an API domain identifying agent.\\[4pt]
\\
Your task:\\
\textbullet\ Evaluate the domain identification's semantic alignment with the API specification (domain, scope, capabilities, standards, ecosystem).\\
\textbullet\ Judge the intrinsic quality of the domain identification itself (clarity, specificity, plausibility/accuracy, relevance; quality of the three ``developments'' bullets), regardless of the spec's completeness.\\
\textbullet\ Provide one final rating (1--5) that balances both alignment and intrinsic quality.\\[4pt]
\\
\textbf{Rating Scale:}\\
\hspace*{8pt}\quad 1: Very Poor --- Major misalignment (wrong or off-scope domain) and low quality. Generic or irrelevant bullets, speculative/incorrect claims, contradictions; missing the required structure; not actionable.\\
\hspace*{8pt}\quad 2: Poor --- Partially relevant but notably flawed. Domain is too broad/narrow or only partly correct; bullets are vague, outdated, or weakly tied to the spec; inconsistencies or superficial content.\\
\hspace*{8pt}\quad 3: Fair --- Mostly aligned with moderate quality. Domain generally correct; bullets somewhat useful but generic or missing specificity/context; minor misinterpretations or gaps; acceptable organization.\\
\hspace*{8pt}\quad 4: Good --- Strong alignment and solid quality. Domain precise and well-scoped; bullets concise, relevant, and plausible with recent context where applicable; clear, actionable takeaways; minor omissions only.\\
\hspace*{8pt}\quad 5: Excellent --- High alignment and high intrinsic quality. Domain is precisely identified with clear boundaries; bullets are specific, current, and materially helpful (e.g., standards/regulatory/architecture trends); organized, balanced, no contradictions; acknowledges uncertainties appropriately.\\[4pt]
\\
\textbf{Output Format (must follow exactly):}\\
Evaluation: (your comments and arguments for the rating)\\
Total rating: (a number between 1 and 5)\\[4pt]
\\
\textbf{Inputs:}\\
API Specification: \{api\_spec\_content\}\\
Domain Identification Output: \{domain\_identification\_content\}\\[4pt]
\\
\textbf{Return Only:}\\
Evaluation: ...\\
Total rating: ...
\end{tcolorbox}

\subsection{Evaluation on Market Research}
\begin{tcolorbox}[colback=blue!3, colframe=blue!50, title=Market Research Evaluation --- LLM-as-a-Judge Prompt, fonttitle=\bfseries\small, boxrule=0.4pt, left=4pt, right=4pt, top=4pt, bottom=4pt, breakable]
\small
\ttfamily
You are an expert evaluator for API market research tasks.\\[4pt]
\\
\textbf{Your Objective:}\\
You will be given two inputs:\\
\textbullet\ An original API specification.\\
\textbullet\ A market summary produced by a web-enabled market search agent.\\[4pt]
\\
Your task:\\
\textbullet\ Evaluate the market summary's semantic alignment with the API specification (scope, domain, capabilities, target users, ecosystem).\\
\textbullet\ Judge the intrinsic quality of the market summary itself (accuracy, sourcing, coverage, structure, clarity, relevance, currency), regardless of the API specification's completeness.\\
\textbullet\ Provide one final rating (1--5) that balances both alignment and intrinsic quality.\\[4pt]
\\
\textbf{Rating Scale:}\\
\hspace*{8pt}\quad 1: Very Poor --- Major misalignment and low quality. Off-scope, factual errors, speculative claims, few/no credible sources, missing core market elements; disorganized and not actionable.\\
\hspace*{8pt}\quad 2: Poor --- Partial relevance with notable gaps. Some correct points but inconsistent focus, limited/weak citations, outdated information, superficial coverage of competitors/pricing/adoption; unclear or contradictory in places.\\
\hspace*{8pt}\quad 3: Fair --- Mostly aligned with moderate quality. Generally accurate, some citations and useful insights, but gaps in coverage (e.g., competitors, pricing, adoption, trends) or uneven sourcing/currency; organization acceptable but needs improvement.\\
\hspace*{8pt}\quad 4: Good --- Strong alignment and solid quality. Well-structured, relevant, mostly comprehensive with credible, recent sources; clear, concise synthesis and actionable takeaways; minor omissions or sourcing/clarity issues.\\
\hspace*{8pt}\quad 5: Excellent --- High alignment and high intrinsic quality. Comprehensive, current, well-cited, and balanced; clearly connects market insights to the API's scope; precise, organized, actionable, no contradictions; helpful, on-scope additions beyond the specification.\\[4pt]
\\
\textbf{Output Format (must follow exactly):}\\
Evaluation: (your comments and arguments for the rating)\\
Total rating: (a number between 1 and 5)\\[4pt]
\\
\textbf{Inputs:}\\
API Specification: \{api\_spec\_content\}\\
Market Summary Output: \{market\_summary\_content\}\\[4pt]
\\
\textbf{Return Only:}\\
Evaluation: ...\\
Total rating: ...
\end{tcolorbox}

\subsection{Evaluation on Gap Analysis}
\begin{tcolorbox}[colback=blue!3, colframe=blue!50, title=Gap Analysis Evaluation --- LLM-as-a-Judge Prompt, fonttitle=\bfseries\small, boxrule=0.4pt, left=4pt, right=4pt, top=4pt, bottom=4pt, breakable]
\small
\ttfamily
You are an expert evaluator for API gap analysis tasks.\\[4pt]
\\
\textbf{Your Objective:}\\
You will be given three inputs:\\
\textbullet\ An original API specification.\\
\textbullet\ Market analysis results identifying leaders and benchmarks for the relevant domain(s).\\
\textbullet\ A gap analysis output produced by a Gap Analysis agent.\\[4pt]
\\
Your task:\\
\textbullet\ Evaluate the gap analysis output's semantic alignment with the API specification and the market analysis results (domain, scope, capabilities, standards, ecosystem, leader comparisons).\\
\textbullet\ Judge the intrinsic quality of the gap analysis itself (coverage of required categories, OpenAPI mapping accuracy, prioritization rigor, clarity, structure, constraints adherence, actionability), regardless of the API specification's completeness.\\
\textbullet\ Provide one final rating (1--5) that balances both alignment and intrinsic quality.\\[4pt]
\\
\textbf{Rating Scale:}\\
\hspace*{8pt}\quad 1: Very Poor --- Major misalignment and low quality. Wrong/unclear domain; ignores or contradicts the specification and market analysis; fails required sections; proposes endpoints; violates constraints (not bullets, wrong length, process/tool notes); generic or unusable; lacks prioritization/rationale.\\
\hspace*{8pt}\quad 2: Poor --- Partially relevant with notable gaps. Some correct observations but inconsistent with the specification or market analysis; missing several required categories or OpenAPI mapping; superficial, vague, or generic; limited prioritization rationale; structural/length violations.\\
\hspace*{8pt}\quad 3: Fair --- Mostly aligned with moderate quality. Covers most required categories; mapping to the specification is present but shallow or with minor inaccuracies; prioritization exists but limited rationale/dependencies; some assumptions unlabeled; minor formatting or length issues; moderate actionability.\\
\hspace*{8pt}\quad 4: Good --- Strong alignment and solid quality. Correct domain inferred from the specification; addresses all required categories; accurate, specific mapping to key paths, models, naming/versioning; clear High/Medium/Low priorities with rationale and constraints; adheres to bullets and 300--500 words; no endpoint proposals; minor omissions only; market comparisons consistent with provided leaders.\\
\hspace*{8pt}\quad 5: Excellent --- High alignment and high intrinsic quality. Precise domain with 1--2 synonyms; thorough and accurate OpenAPI mapping (paths, models, naming/versioning, coverage gaps, breaking-change risks); category-by-category gaps vs market-leading expectations grounded in the specification and market analysis; well-justified prioritization with dependencies/constraints and labeled assumptions; concise bullet structure within 300--500 words; focused ``Notes for design'' aligned to naming/versioning, async/webhooks, and compliance; no contradictions or scope drift.\\[4pt]
\\
\textbf{Output Format (must follow exactly):}\\
Evaluation: (your comments and arguments for the rating)\\
Total rating: (a number between 1 and 5)\\[4pt]
\\
\textbf{Inputs:}\\
API Specification: \{api\_spec\_content\}\\
Market Analysis Results: \{market\_analysis\_content\}\\
Gap Analysis Output: \{gap\_analysis\_content\}\\[4pt]
\\
\textbf{Return Only:}\\
Evaluation: ...\\
Total rating: ...
\end{tcolorbox}

\subsection{Evaluation on Endpoint Suggestions}
\begin{tcolorbox}[colback=blue!3, colframe=blue!50, title=Endpoint Suggestions Evaluation --- LLM-as-a-Judge Prompt, fonttitle=\bfseries\small, boxrule=0.4pt, left=4pt, right=4pt, top=4pt, bottom=4pt, breakable]
\small
\ttfamily
You are an expert evaluator for API endpoint design and enhancement tasks.\\[4pt]
\\
\textbf{Your Objective:}\\
You will be given two or more inputs:\\
\textbullet\ An OpenAPI specification for the product.\\
\textbullet\ An endpoint plan produced by an EndpointPlannerAgent (suggested endpoints and specification modifications).\\
\textbullet\ Optional: Market Summary Report and/or Gap Analysis Report that prioritized market gaps to address.\\[4pt]
\\
Your task:\\
\textbullet\ Evaluate the endpoint plan's semantic alignment with the OpenAPI specification and the provided market or gap analysis inputs (coverage of prioritized gaps, consistency with existing API patterns).\\
\textbullet\ Judge the intrinsic quality of the endpoint plan itself (API best practices, clarity, testability, security and compliance, performance and SLA considerations, backward compatibility, and format adherence), regardless of the completeness of the provided inputs.\\
\textbullet\ Provide one final rating (1--5) that balances both alignment and intrinsic quality.\\[4pt]
\\
\textbf{Rating Scale:}\\
\hspace*{8pt}\quad 1: Very Poor --- Major misalignment and low quality. Endpoints conflict with existing API patterns, ignore prioritized gaps, or violate required constraints; mini-specifications are missing; change types are unclear; security and performance considerations are absent; output is not actionable.\\
\hspace*{8pt}\quad 2: Poor --- Partial alignment with notable gaps. Some viable ideas exist but endpoint designs are inconsistent with naming conventions, versioning, authentication, pagination, or filtering patterns; weak traceability to identified gaps; specifications are vague or difficult to validate; migration guidance is missing; formatting or length requirements are not followed.\\
\hspace*{8pt}\quad 3: Fair --- Mostly aligned with moderate quality. Suggested endpoints are generally viable but contain ambiguities or omissions such as idempotency requirements, detailed error models, or performance targets; minor deviations from established API conventions; uneven coverage of prioritized gaps; useful but requires refinement.\\
\hspace*{8pt}\quad 4: Good --- Strong alignment and solid quality. Endpoint proposals follow existing API patterns, address most High and Medium priority gaps, provide clear mini-specifications, and include appropriate security and performance considerations; changes are additive unless a justified breaking change is identified; only minor improvements are needed.\\
\hspace*{8pt}\quad 5: Excellent --- High alignment and high intrinsic quality. Endpoint proposals comprehensively address prioritized gaps with precise and testable mini-specifications; fully consistent with existing API conventions and design patterns; clear treatment of security, compliance, SLAs, pagination, filtering, error handling, and idempotency; explicit change types and migration guidance; concise, well-structured, and highly actionable.\\[4pt]
\\
\textbf{Output Format (must follow exactly):}\\
Evaluation: (your comments and arguments for the rating)\\
Total rating: (a number between 1 and 5)\\[4pt]
\\
\textbf{Inputs:}\\
OpenAPI Specification: \{openapi\_spec\_content\}\\
Endpoint Suggestions Output: \{endpoint\_suggestions\}\\[4pt]
\\
\textbf{Return Only:}\\
Evaluation: ...\\
Total rating: ...
\end{tcolorbox}

\begin{tcolorbox}[colback=gray!10,colframe=gray!50,boxrule=0.5pt]
\textbf{Note:} For a detailed case study, please refer to the appendix sections below. 
\end{tcolorbox}
\onecolumn
\section{Case Study}
\label{app:case_study}
In this section, we present a case study of Agent4Spec using Claude-Sonnet-4.6.
The transcript below shows the exact step-by-step sequential multi-agent execution timeline, functional tool invocations, and full response texts logged during runtime.
A video recording the entire process with human interaction is provided in the supplementary materials.

\begin{userbox}{Event 1}
hi
\end{userbox}

\begin{agentbox}{OrchestratorAgent}
Hello! Would you like to use our multi-agent system to assess, validate, or improve an API specification? If so, please share the file path to your API specification (e.g., \texttt{/path/to/api.yaml}) and let me know how you'd like to proceed.
\end{agentbox}

\begin{userbox}{Event 2}
\texttt{/Users/Handsome/Desktop/code/api\_specs/api\_norm\_service.yaml}
\end{userbox}

\begin{agentbox}{OrchestratorAgent}
Thank you for providing the file path:\\ \texttt{/Users/Handsome/Desktop/code/api\_specs/api\_norm\_service.yaml}. \\
How would you like to proceed? Here are your options:
\begin{enumerate}
    \item \textbf{Run Syntax Validation}: Check for any errors or inconsistencies...
    \item \textbf{Compare Against Market Leaders}: Analyze the API against top competitors...
    \item \textbf{Revise the Specification}: Based on validation or market analysis.
\end{enumerate}
\end{agentbox}

\begin{userbox}{Event 3}
Run Syntax Validation
\end{userbox}

\begin{systembox}{System / Tool Usage}
\begin{itemize}
    \item \texttt{OrchestratorAgent} calls \texttt{transfer\_to\_agent(agent\_name="SyntaxCheckAgent")}.
    \item \texttt{SyntaxCheckAgent} calls \texttt{read\_yaml\_file(file\_path=".../api\_norm\_service.yaml")} and \texttt{validate\_openapi\_detailed(spec\_path="...")}.
    \item \textbf{Tool Response}: Validation fails. Identified an \texttt{OpenAPIValidationError} on instance \texttt{['info']}: the property \texttt{descriptio} does not match regexes (typo for \texttt{description}).
\end{itemize}
\end{systembox}

\begin{agentbox}{SyntaxCheckAgent}
\textbf{Summary}
\begin{itemize}[noitemsep]
    \item \textbf{File Type}: YAML
    \item \textbf{Detected OAS Version}: 3.0.1
    \item \textbf{Overall Status}: Invalid OpenAPI Specification
\end{itemize}

\textbf{Issues}
\begin{enumerate}[noitemsep]
    \item \textbf{Location}: \texttt{info} block.
    \item \textbf{Error}: The property \texttt{descriptio} does not match the expected key \texttt{description}.
    \item \textbf{Cause}: Typo in the attribute name \texttt{descriptio} (should be \texttt{description}).
\end{enumerate}

\textit{[Agent provides the corrected YAML snippet replacing 'descriptio' with 'description']}

Would you like me to forward this corrected version to the \texttt{[SpecificationGenerationAgent4SyntaxValidation]} agent to generate the complete and validated specification?
\end{agentbox}

\begin{userbox}{Event 4}
forward to the [SpecificationGenerationAgent4SyntaxValidation] agent to generate
\end{userbox}

\begin{systembox}{System / Tool Usage}
\begin{itemize}
    \item \texttt{SyntaxCheckAgent} calls \texttt{transfer\_to\_agent}.
    \item \texttt{SpecificationGenerationAgent4SyntaxValidation} calls \texttt{read\_yaml\_file(...)}.
    \item \texttt{SpecificationGenerationAgent4SyntaxValidation} calls \texttt{save\_openapi\_spec(...)} with the corrected YAML string to output file \texttt{api\_norm\_service\_corrected.yaml}.
    \item \textbf{Tool Response}: \texttt{\{"Status": "Successfully saved"\}}.
    \item \texttt{SpecificationGenerationAgent4SyntaxValidation} calls \texttt{transfer\_to\_agent(\allowbreak agent\_name="OrchestratorAgent")}.
\end{itemize}
\end{systembox}

\begin{agentbox}{OrchestratorAgent}
The corrected API specification has been successfully saved at the following path:\\
\texttt{/Users/Handsome/Desktop/code/api\_specs/api\_norm\_service\_corrected.yaml}

Would you like to proceed with any of the following steps?
\begin{itemize}[noitemsep]
    \item Market Comparison
    \item Re-validate the Corrected Specification
    \item Further Revise the Specification
    \item End the session.
\end{itemize}
\end{agentbox}

\begin{userbox}{Event 5}
Re-validate the Corrected Specification: Ensure the updated version is now fully valid.
\end{userbox}

\begin{systembox}{System / Tool Usage}
\begin{itemize}
    \item \texttt{OrchestratorAgent} calls \texttt{transfer\_to\_agent(agent\_name="SyntaxCheckAgent")}.
    \item \texttt{SyntaxCheckAgent} calls \texttt{validate\_openapi\_detailed(spec\_path=\\".../api\_norm\_service\_corrected.yaml")}.
    \item \textbf{Tool Response}: \texttt{\{"Is\_valid: ": true, "Errors": "Specification is valid!"\}}.
\end{itemize}
\end{systembox}

\begin{agentbox}{SyntaxCheckAgent}
\textbf{Validation Result:}
\begin{itemize}[noitemsep]
    \item \textbf{File Path}: \texttt{.../api\_specs/api\_norm\_service\_corrected.yaml}
    \item \textbf{Status}: Valid OpenAPI Specification
    \item \textbf{Detected Issues}: None
\end{itemize}
Your corrected specification is now fully validated and adheres to OpenAPI standards. It is ready for use or further processing.
\end{agentbox}

\begin{userbox}{Event 6}
Market Comparison: Analyze the specification against competitors, identify gaps, and get improvement suggestions.
\end{userbox}

\begin{systembox}{System / Tool Usage}
\begin{itemize}
    \item \texttt{SyntaxCheckAgent} calls \texttt{transfer\_to\_agent(agent\_name=\\"SequentialMarketSearchAgent")}.
\end{itemize}
\end{systembox}

\begin{agentbox}{DomainIdentificationAgent}
Let me define the domain and search for relevant market intelligence.\\
\textbf{Domain:} Product / Part Number Validation \& Normalization (Industrial Manufacturing PLM)\\
\textbf{Synonyms \& Related Terms:}
\begin{itemize}
    \item Product Identifier Validation API
    \item PLM Part Number Normalization Service
    \item GTIN / MLFB / Industrial Numbering Standards API
\end{itemize}
\end{agentbox}

\begin{systembox}{System / Tool Usage}
\begin{itemize}
    \item \texttt{MarketResearcherAgent} calls \texttt{search\_and\_fetch\_content\_free} for queries: ``NORM service API Siemens numbering standards competitors'' and ``GTIN MLFB validation API pricing and market adoption''.
    \item \textbf{Tool Response}: Successfully fetches competitor pages (e.g., Siemens API Catalog, ETSI standards, EAN Search API, GS1 standards).
\end{itemize}
\end{systembox}

\begin{agentbox}{MarketResearcherAgent}
\textbf{Market Research Summary}: Product Number Validation \& Normalization in Industrial Manufacturing PLM

\textbf{Concise Summary (Market Overview)}
The Siemens NORM Service operates in the niche intersection of Product Data Quality and Industrial PLM Microservices, providing validation and normalization of industrial product identifiers (MLFB, GTIN, SSN, TNS, SFC, SHS). This market sits within the broader Data Quality Tools sector, valued at USD 2.71 billion in 2024 and projected to reach USD 4.15 billion by 2032 (CAGR 5.46\%), while the wider Enterprise Data Quality Services market is estimated at USD 10 billion in 2024, targeting USD 25 billion by 2033 (CAGR 10\%).

\textbf{Key players in this space include}:
\begin{itemize}[noitemsep]
    \item \textbf{GS1}: With its ``Verified by GS1'' service for global GTIN/EAN validation and trusted product data exchange.
    \item \textbf{EAN-Search.org}: Offering a public barcode lookup API covering over 1.2 billion EAN/GTIN records with REST and GraphQL endpoints.
    \item \textbf{Microsoft Dynamics 365}: Which embeds product identifier management (GTIN, UPC, etc.) within its ERP/SCM platform.
\end{itemize}
Within Siemens' own ecosystem, the Siemens Developer Portal (\texttt{developer.siemens.com}) exposes 100+ industrial APIs including the Product Information Hub for standardized product data, and the Drivetrain Analyzer Cloud which references MLFB product numbers as core asset identifiers---confirming MLFB normalization is foundational infrastructure for Siemens' digital twin and Industrial IoT strategy.

The market is driven by regulatory pressures (EU Digital Product Passport, GS1 standards), the shift from monolithic PLM to cloud-native microservices, and the growing dependency of AI/ML analytics on clean, standardized product data. GS1 Germany charges a one-time service price plus recurring annual usage fees scaled by company turnover for GTIN numbering capacity, indicating a usage-based monetization model is standard.

\textbf{Searched Websites}
\begin{enumerate}[noitemsep]
    \item \texttt{https://www.ean-search.org/}
    \item \texttt{https://www.gs1.org/services/verified-by-gs1}
    \item \texttt{https://www.gs1india.org/}
    \item \texttt{https://ref.gs1.org/architecture/system-architecture/}
    \item \texttt{https://www.gs1-germany.de/en/prices/}
    \item \texttt{https://developer.siemens.com/apis.html}
\end{enumerate}
\end{agentbox}

\begin{agentbox}{GapIdentifierAgent}
\textbf{Gap Analysis --- NORM Service v3.5.0}

\textbf{Domain:} Product Number Validation \& Normalization (Industrial Manufacturing PLM) --- synonyms: Part Number Standardization API, Product Identifier Normalization Service.

\textbf{Gap Summary}
\begin{itemize}[noitemsep]
    \item \textbf{Features / Coverage (Priority: High)}: Single operation (\texttt{POST /validate}) only; no individual-system endpoints, no lookup/search, no conversion between numbering systems, no checksum verification endpoint. Leaders (GS1 Verified, EAN-Search) offer lookup, search, similar-product-search, and bulk export.
    \item \textbf{Features / Coverage (Priority: Medium)}: No support for emerging identifiers: GS1 DataMatrix/2D barcodes, EU Digital Product Passport identifiers, IEC 61406 Identification Link. Market is moving toward these.
    \item \textbf{Security / Compliance (Priority: Medium)}: Bearer JWT (Entra ID) only; no API-key alternative, no OAuth2 scopes/RBAC granularity, no documented audit/logging of validation requests. GS1 and Dynamics 365 expose fine-grained RBAC and audit trails.
    \item \textbf{Developer Experience / Docs / SDKs (Priority: High)}: No SDK, no client libraries, no OpenAPI description (key misspelled as \texttt{descriptio}), no license field, no error-response schemas for 4xx/5xx, no health/status endpoint. EAN-Search provides libraries for 10+ languages and GraphQL.
    \item \textbf{Ecosystem / Global Reach (Priority: Medium)}: Single \texttt{eu-central-1} AWS Lambda behind one APIM gateway; no multi-region failover. Siemens Insights Hub APIs expose multi-region endpoints. No webhook/event integration for downstream consumers (e.g., digital twin pipelines).
\end{itemize}

\textbf{Existing Resources \& Patterns}
\begin{itemize}[noitemsep]
    \item \textbf{Key path}: \texttt{POST /norm/v3/validate} --- sole endpoint; accepts JSON array of strings (max 1,000), returns \texttt{NumberDto[]}.
    \item \textbf{Naming / Versioning}: Path-based versioning (\texttt{/v3}); camelCase fields; tag label ``v3'' only. No v1/v2 paths or deprecation markers.
    \item \textbf{Coverage gaps}: Only 200 response defined --- no 400 (malformed JSON), 401, 403, 413 (oversize batch), 429, 500 schemas. \texttt{AbstractNumber} is an empty object --- adds no value in the \texttt{allOf} composition.
    \item \textbf{Breaking-change risks}: The \texttt{oneOf} discriminator is implicit (no discriminator property set) --- adding a new numbering system could confuse code-gen clients. Field \texttt{descriptio} (typo) may break doc generators. \texttt{x-aws-lambda} is a non-standard extension tied to infra.
\end{itemize}

\textbf{Prioritized Gaps Summary}
\begin{itemize}[noitemsep]
    \item \textbf{High}: Batch / async processing \& pagination; Error-response schemas (4xx/5xx); Developer experience (SDKs, health endpoint, spec hygiene).
    \item \textbf{Medium}: Emerging identifier support (GS1 DataMatrix, Digital Product Passport); Security granularity (scopes, RBAC); Multi-region / failover infrastructure.
    \item \textbf{Low}: Metering / quota headers for internal chargeback.
\end{itemize}
\end{agentbox}

\begin{agentbox}{EndpointPlannerAgent}
\textbf{Endpoint Plan --- NORM Service v3.5.0}

\textbf{Suggested Endpoints}
\begin{center}
\small
\begin{tabularx}{\linewidth}{@{}l p{4.5cm} X l l@{}}
\toprule
\textbf{Method} & \textbf{Path} & \textbf{Purpose} & \textbf{Priority} & \textbf{Type} \\
\midrule
\texttt{POST} & \texttt{/norm/v3/validate/batch} & Async bulk validation ($>1,000$ numbers) & High & Additive \\
\texttt{GET} & \texttt{/norm/v3/jobs/\{jobId\}} & Poll async job status/results & High & Additive \\
\texttt{GET} & \texttt{/norm/v3/health} & Health/readiness check & High & Additive \\
\texttt{GET} & \texttt{/norm/v3/systems} & List supported numbering systems & High & Additive \\
\texttt{POST} & \texttt{/norm/v3/validate} & Add 4xx/5xx error response schemas & High & Mod \\
\texttt{POST} & \texttt{/norm/v3/convert} & Convert between numbering formats & Medium & Additive \\
\texttt{POST} & \texttt{/norm/v3/identify} & Auto-detect numbering system type & Medium & Additive \\
\bottomrule
\end{tabularx}
\end{center}

\textbf{Endpoint Mini-Specs}

\textbf{1. POST \texttt{/norm/v3/validate/batch} --- Async Bulk Validation}
\begin{itemize}[noitemsep]
    \item \textbf{Request Headers}: \texttt{Authorization: Bearer \{JWT\}}, \texttt{Content-Type: application/json}, \texttt{Idempotency-Key: \{uuid\}} (Required).
    \item \textbf{Body Layout}: \texttt{\{ "numbers": string[], "callbackUrl"?: string \}} (Max 100,000 items).
    \item \textbf{Responses}:
    \begin{itemize}
        \item \texttt{202 Accepted}: \texttt{\{ "jobId": "uuid", "status": "queued", "estimatedCompletionSeconds": 30 \}}
        \item \texttt{400 Bad Request}, \texttt{413 Payload Too Large} (Exceeds 100,000), \texttt{429 Rate Limited}.
    \end{itemize}
    \item \textbf{Rate Limits}: 10 req/min per client; exposes standard \texttt{X-RateLimit} tracking headers.
    \item \textbf{Webhook Event}: Fires \texttt{norm.batch.completed} or \texttt{norm.batch.failed}. Contains response cryptographic signature validation via \texttt{X-Norm-Signature: HMAC-SHA256(\{shared\_secret\}, body)}.
\end{itemize}

\textbf{2. GET \texttt{/norm/v3/jobs/\{jobId\}} --- Poll Async Job Status}
\begin{itemize}[noitemsep]
    \item \textbf{Parameters}: \texttt{jobId} (Path, UUID), \texttt{offset} (Query, default 0), \texttt{limit} (Query, default 100).
    \item \textbf{Responses}: \texttt{200 OK} returns status payload (\texttt{"queued" | "processing" | "completed" | "failed"}) along with progress integers and paginated \texttt{NumberDto[]} items. Supports caching via \texttt{ETag}.
\end{itemize}

\textbf{3. Modifications to Existing POST \texttt{/norm/v3/validate}}
\begin{itemize}[noitemsep]
    \item Add strict response schemas for error codes \texttt{400}, \texttt{401}, \texttt{403}, \texttt{413}, \texttt{429}, and \texttt{500}.
    \item Inject polymorphic explicit object discriminator to \texttt{NumberDto.output.oneOf}: \texttt{\{ "propertyName": "system" \}}.
\end{itemize}

\textbf{Modifications \& Deprecations Plan}
\begin{itemize}[noitemsep]
    \item \textbf{Fix Typo}: \texttt{info.descriptio} $\rightarrow$ \texttt{info.description} (Flagged for minor version bump).
    \item \textbf{Prune Spec}: Remove empty structural parent object \texttt{AbstractNumber}; replace directly with flat target polymorphism schemas inside the main data array execution block.
    \item \textbf{Infra Decoupling}: Strip non-standard API Gateway vendor extensions (\texttt{x-aws-lambda}) from the public interface distribution asset file.
\end{itemize}

\textbf{Assumptions \& Constraints}
\begin{itemize}[noitemsep]
    \item Siemens Entra ID can be extended with custom OAuth2 scopes (\texttt{norm.validate.write}, \texttt{norm.jobs.read}).
    \item Idempotency-Key and async job patterns successfully align with Siemens Insights Hub engineering precedents (\textit{Agent Management Async API}).
    \item Job result retention of 7 days matches internal enterprise transient data compliance limitations.
\end{itemize}
\end{agentbox}

\begin{agentbox}{OrchestratorAgent}
\textbf{What would you like to do next?}
\begin{enumerate}
    \item \textbf{Generate revised specification} --- I can call the \texttt{SpecificationGenerationAgent} to produce an updated OpenAPI spec incorporating these endpoint proposals.
    \item \textbf{Continue market comparison} --- Further analyze competitive gaps or drill into specific areas.
\end{enumerate}
Which would you prefer?
\end{agentbox}

\begin{userbox}{Event 7}
Wrap up the session
\end{userbox}

\begin{agentbox}{OrchestratorAgent}
\textbf{Session Complete}

All market comparison data --- domain summary, competitor analysis, gap report, and suggested endpoints --- has been captured.

\textbf{If you'd like to come back and take further action, here are your options:}
\begin{enumerate}[noitemsep]
    \item Run syntax validation on your spec
    \item Generate a revised specification incorporating the market-driven endpoint suggestions above.
    \item Re-run market comparison with different domain hints or constraints
\end{enumerate}
\textbf{Thanks for using the API Analysis Assistant! Feel free to return anytime.}
\end{agentbox}

\end{document}